\documentclass[%
 aip,
 amsmath,amssymb,
 reprint,%
]{revtex4-1}

\usepackage{graphicx}
\usepackage{dcolumn}
\usepackage{bm}
\usepackage{booktabs}
\usepackage[utf8]{inputenc}
\usepackage[T1]{fontenc}
\usepackage{mathptmx}
\usepackage{etoolbox}
\usepackage{xcolor}

\usepackage{hyperref}
\hypersetup{
    colorlinks=true,
    linkcolor=blue,
    filecolor=magenta,      
    urlcolor=blue,
    citecolor=blue,
    pdftitle={Overleaf Example},
    pdfpagemode=FullScreen,
    }

\makeatletter
\def\@email#1#2{%
 \endgroup
 \patchcmd{\titleblock@produce}
  {\frontmatter@RRAPformat}
  {\frontmatter@RRAPformat{\produce@RRAP{*#1\href{mailto:#2}{#2}}}\frontmatter@RRAPformat}
  {}{}
}%
\makeatother
\begin{document}

\title{Extending the limits of rotationally resolved mid-IR spectroscopy: an accurate model of trioxane derived from combining Fourier transform and sub-Doppler measurements}
\author{Pierre Asselin}
\email{pierre.asselin@sorbonne-universite.fr}
\affiliation{Sorbonne Université, CNRS, MONARIS, UMR 8233, Paris, France}
\author{Dang-Bao-An Tran}%
\altaffiliation{Current address: Time and Frequency Department, National Physical Laboratory, Teddington, United Kingdom}
\affiliation{Université Sorbonne Paris Nord, Laboratoire de Physique des Lasers, LPL, CNRS, UMR 7538, F‐93430, Villetaneuse, France}
\affiliation{Faculty of Physics, Ho Chi Minh City University of Education, Ho Chi Minh City, Vietnam}

\author{Alexander Shelkovnikov}%
\altaffiliation{Permanent address: P.N. Lebedev Physics Institute, Russian Academy of Sciences, Leninsky prosp. 53,
119991 Moscow, Russia
}%
\affiliation{Université Sorbonne Paris Nord, Laboratoire de Physique des Lasers, LPL, CNRS, UMR 7538, F‐93430, Villetaneuse, France}

\author{Sean Tokunaga}%
\affiliation{Université Sorbonne Paris Nord, Laboratoire de Physique des Lasers, LPL, CNRS, UMR 7538, F‐93430, Villetaneuse, France}%

\author{Mathieu Manceau}%
\affiliation{Université Sorbonne Paris Nord, Laboratoire de Physique des Lasers, LPL, CNRS, UMR 7538, F‐93430, Villetaneuse, France}%

\author{Benoît Darquié}
\email{benoit.darquie@univ-paris13.fr}
\affiliation{Université Sorbonne Paris Nord, Laboratoire de Physique des Lasers, LPL, CNRS, UMR 7538, F‐93430, Villetaneuse, France}

\begin{abstract}
1,3,5-trioxane, a polyoxymethylene (POM) molecule relevant to prebiotic chemistry and astrophysics, exhibits a dense rovibrational spectrum that challenges high-resolution spectroscopic characterization. We report an extensive study of its $\nu_{5}$ CO symmetric stretching mode (960–1000 cm$^{-1}$) by combining broadband high-resolution FTIR spectroscopy (0.001 cm$^{-1}$ resolution, $>$2000 transitions) with SI-traceable sub-Doppler saturated absorption measurements using frequency-stabilized CO$_{2}$ lasers and a frequency-comb-referenced quantum cascade laser. The latter achieve line-position uncertainties as low as 40 kHz, revealing inaccuracies in the FTIR frequency scale. A frequency-scale correction is determined by maximizing the global consistency of the complete spectroscopic data set. A global fit combining the calibrated FTIR data, the SI-traceable sub-Doppler measurements and previously reported millimeter-wave rotational transitions yields a refined spectroscopic model. The resulting analysis significantly improves the precision of the rotational, quartic, and sextic constants for both the ground and $\nu_{5}$ excited vibrational states. This work extends mid-infrared frequency metrology to the largest molecule studied to date (12 atoms), providing a dense set of reference transitions for frequency calibration around 970 cm$^{-1}$. The resulting spectroscopic parameters reach relative uncertainties as low as $3.7\times 10^{-10}$ for the band origin, making them valuable for molecular databases used in astrophysics and atmospheric physics, as well as for precision measurements in fundamental physics. Additionally, the observed line shapes indicate the absence of significant intramolecular vibrational energy redistribution (IVR) within the investigated energy range, while a few discrepancies in the Hamiltonian fit suggest the onset of weak perturbations, demonstrating the sensitivity of the present approach to probe subtle intramolecular couplings in complex polyatomic spectra.
\end{abstract}

\maketitle

\section{\label{sec_1}Introduction}

The trimer of formaldehyde, 1,3,5-trioxane or (H$_2$CO)$_3$, is of the polyoxymethylene (POM) class of molecules, considered highly relevant to prebiotic chemistry as a source of formaldehyde in cometary comae \cite{cottin2000experimental,cottin2004origin}. However, no conclusive astronomical detection of POMs has been reported to date. 1,3,5-trioxane’s dense rotational spectrum resulting from its relatively large number of atoms and high mass could serve as suitable radioastronomy tracer. With its rich microwave, THz and mid-infrared (IR) spectrum, trioxane is also a promising molecule for the realization of optically-pumped far-IR gas lasers \cite{dangoisse1981assignment} (for which there is recently renewed interest \cite{pagies2016low,amirzhan2022quantum}) and for frequency calibration and metrology with its almost continuous set of transitions (see below) that could be used as a grid of references in the 850-1500 cm$^{-1}$ spectral window (five times more than the usual OsO$_{4}$ \cite{acef1999accurate},  SF$_{6}$ \cite{acef2000new}, and CO$_{2}$ \cite{amy2004absolute} absolute frequency grids).

The first trioxane microwave spectrum was recorded by Oka \textit{et al} \cite{oka1964microwave}, followed by a series of sub-millimeter wave (sub-mmw) rotational studies by Colmont \textit{et al} \cite{colmont1975microwave} in the ground and few low-lying vibrationally excited states $\nu_7$, $\nu_{19}$, $\nu_{20}$, $2\nu_{20}$, $\nu_{7}+\nu_{20}$, and $\nu_{19}+\nu_{20}$ up to 850 cm$^{-1}$ above the ground state. These rotational analyses provided first reliable determinations of ground and excited state molecular parameters. Note also the pioneering pump-probe mmw-infrared (IR) laser experiments by Dangoisse \textit{et al} \cite{dangoisse1981assignment} (based on a far-IR trioxane laser, optically pumped by a CO$_{2}$ laser) in which 36 rotational transitions within the $\nu_{5}$ state were assigned. In the mid-IR range, two jet-cooled studies reported rovibrational analyses of $\nu_{17}$ and $\nu_{16}$ bands centred at 1071 and 1177~cm$^{-1}$, respectively \cite{henninot1992infrared,gibson2015rotationally}. Recently, a far-IR high resolution (HR) Fourier Transform (FT) study using a long pass cell and the AILES beamline of the SOLEIL synchrotron yielded rovibrational spectra of the three lowest fundamental bands $\nu_7$, $\nu_{19}$, and $\nu_{20}$ of 1,3,5-trioxane below 650 cm$^{-1}$ \cite{richard2022high}.  Finally, we must cite our own latest sub-Doppler study of $\nu_{5}$ fundamental band \cite{tran2025extending}, the data of which, combined with newly reported materials, are further analyzed in the present work.

Accurate models of such large polyatomic molecules are scarce. This is because producing such a model requires, as reported in this work, a consorted effort to study the species using suitably high-resolution spectroscopic methods over wide regions, and an associated analytical procedure to transform all the data into a useful model.

In addition, precision studies of rovibrational level structures of such molecules are particularly rare at mid-IR energies, since many of the highest resolution spectroscopic techniques (microwave or sub-mmw) do not contain any signal from excited vibrational states, often unpopulated at room temperature. And direct mid-IR probes of rovibrational transitions are often either broad but insufficiently resolved, or vice-versa.

The solution proposed in this work is to jointly analyze two such complementary mid-IR datasets: broadband (40 cm$^{-1}$ span) conventional FT measurements at a 30 MHz full width at half maximum (FWHM) resolution (the highest commercially available) and narrower band ($\sim$10 GHz span) saturation spectroscopy data at a much higher resolution ($\sim$400~kHz). Once combined, these allow us to determine molecular parameters with greatly improved accuracy compared to those derived by exploiting them separately.

We emphasize that, together with our previous accompanying paper devoted to saturated absorption spectroscopy of trioxane, \cite{tran2025extending} the present work extends mid-IR sub-Doppler spectroscopy to the largest molecule (trioxane, (H$_2$CO)$_3$ has 12 atoms) investigated to date by this technique. We report absolute frequencies for 557 rovibrational transitions at resolutions and uncertainties as low as $\sim$400 kHz and $\sim$40 kHz respectively, which is unprecedented for a species of this complexity. Beyond providing a dense set of highly accurate frequency references, these measurements reveal inaccuracies in the FTIR frequency scale at the megahertz level. Rather than relying on a conventional calibration based on a limited number of reference lines, we determine the correction to apply to the FTIR frequency-scale by optimizing the agreement of the complete data set – including the corrected FTIR transitions, the sub-Doppler measurements, and previously reported millimetre-wave data – with a global rovibrational model. This model-assisted post-calibration is a central element of the methodology developed here, enabling the SI-traceable accuracy of the sub-Doppler measurements to be propagated to the broadband FTIR data prior to the final determination of the molecular Hamiltonian. The resulting analysis yields a significantly improved rovibrational model of the $\nu_5$ parallel band (A$_{1}$ symmetry CO stretching mode) of 1,3,5-trioxane in the 960-1000 cm$^{-1}$ region.

This confirms the potential of the technologies and procedures we develop for bringing increasingly complex species within reach of ultra-precise measurements. This has far reaching applications from Earth sciences and astrophysics, to metrology, by enabling molecular databases to be populated with increasingly accurate parameters \cite{polyansky2015high,vaskuri2018uncertainty,roueff2020spectroscopic}. Quantum state resolved mid-IR spectroscopy and the precise measurement of transition energies between molecular eigenstates in increasingly complex species is also required for exploiting molecules in quantum technologies \cite{schulz2015many,liu2023ergodicity,tesch2002quantum,albert2024quantum} and as a crucial groundwork for conducting fundamental physics measurements \cite{shelkovnikov2008stability,jansen2014perspective,muller2021study,fiechter2022toward,gaul2020chiral}. 

\section{Experimental details} \label{sec:2}
Trioxane is a large molecule with several advantages to realize precision measurements in the mid-IR window: from a spectroscopic point of view, it is a rather rigid oblate top-symmetric (C$_{3v}$) molecule with three low-lying vibrational modes below 600 cm$^{-1}$ of which the lowest falls around 300 cm$^{-1}$, probably resulting in moderate intramolecular vibrational redistribution (IVR) at higher energies. Moreover, gas phase experiments are possible given its relatively high vapour pressure (11 mbar at 293 K \cite{de1983thermodynamic}).

\subsection{Fourier transform infrared spectroscopy}\label{subsec:2_1}

The broadband rovibrational spectrum of 1,3,5-trioxane (>99$\%$ purity, Sigma-Aldrich) has been recorded in the 8.5-12.5 {\textmu}m region at the maximal resolution of 0.001 cm$^{-1}$ using a HR Bruker IFS 125 Fourier Transform InfraRed (FTIR) spectrometer (boxcar apodization mode) located at the AILES beamline of the SOLEIL synchrotron facility. A vapour of 1,3,5-trioxane was injected in a room temperature multipass cell equipped with Au-Cr coated glass mirrors and isolated from the FT spectrometer by ZnSe antireflection coated windows for which the White-type optical arrangement was set to obtain 140 m absorption length. The cell pressure was adjusted to 0.48 Pa to optimize the absorption of the $\nu_5$ fundamental band (calculated intensity: 82 km/mol) without saturating the band. The optimal source/detector/beam-splitter configuration for the FT interferometer in the mid-IR region combines the internal globar continuum source, a KBr beam-splitter and a liquid-nitrogen-cooled mercury cadmium telluride (MCT) photodetector equipped with a pass band filter centered around 1000 cm$^{-1}$ (200 cm$^{-1}$ FWHM). The absolute frequency calibration of the FT spectrometer was evaluated by comparing the frequencies of 39 rovibrational lines recorded in the $\nu_2$ O-C-O bending mode region with those reported in the last update of HITRAN 2020 \cite{gordon2022hitran2020}, leading to a line frequency average deviation equal to 0.00005 cm$^{-1}$ with standard deviation equal to $2.1 \times 10^{-4}$ cm$^{-1}$ (roughly equal to the typical experimental precision for cell-FTIR experiments at 0.001 cm$^{-1}$ resolution). The state-of-the-art multipass cell-HR FTIR set-up available on the AILES beamline fulfills the optimal conditions to resolve the rotational structure of large molecules in the mid-IR range up to high $J$, $K$ quantum number values, even in congested room temperature conditions, while allowing very low-pressure conditions to limit collisional broadening effects. Such benefits turn out to be crucial in the perspective of combining the analysis of rovibrational bands observed in both FTIR and sub-MHz resolution saturated absorption spectra.   

\subsection{SI-traceable ultrahigh-resolution saturated absorption spectroscopy}\label{subsec:2_2}
At Laboratoire de Physique des Lasers, we have conducted sub-Doppler spectroscopy of trioxane in cells at room temperature using two different ultrahigh-resolution metrology-grade spectrometers traceable to the international system of units (SI): one combining ultra-stable frequency-stabilized CO$_2$ lasers and a double-pass cell, the other integrating a widely tunable frequency-comb-stabilized-quantum cascade laser (QCL) and a multi-pass cell.

\subsubsection{Frequency-stabilized-CO$_2$-laser-based spectrometer}
Details of this set-up can be found elsewhere \cite{stoeffler2011high}. A schematic overview of the CO$_2$-laser spectrometer is also shown in Fig. S1 of the Supplementary Material. It is based on the combination of two CO$_2$ lasers. The first is locked to an OsO$_4$ rovibrational line and as such constitutes the secondary frequency standard of this spectral region. The other CO$_2$ laser, the beam of which is used to probe trioxane, is phase-locked to the first with a tuneable radio frequency (RF) offset. Using this locking scheme, we know the absolute frequency of the laser to within less than 60 Hz (see Table S1) and obtain a line width of 10 Hz \cite{stoeffler2011high}. Computer control of the synthesizer delivering the RF offset enables continuous tuneability over the laser gain curve width, \textit{i.e}. $\sim$80 MHz. Acousto-optic modulators allow the accessible spectral window to be increased by around $\pm$300 MHz around each CO$_2$ laser emission line.

Saturated absorption spectra of room temperature trioxane have been recorded in a 58-cm long cell ended with anti-reflective coated wedged ZnSe windows. The laser beam is sent to the absorption cell (pump beam), retro-reflected with a mirror (probe beam) and focused on a liquid-nitrogen-cooled MCT photodetector. A pump-to-probe-beams power ratio of 10 is obtained by placing optical attenuators between the retro-reflecting mirror and the cell. The beam waist inside the cell is on the order of 5 mm. Spectra are recorded using frequency modulation (FM) of the probing CO$_2$ laser at 5015 Hz, with a FM amplitude of 200 kHz, applied on the piezoelectric transducers controlling the laser cavity length. Second-harmonic detection is used so as to flatten the baseline induced by the laser gain curve.

Saturated absorption spectra have been recorded around the $R$(6), $R$(8), $R$(14), $R$(16), $R$(22), and $R$(24) CO$_2$ laser emission lines, with the first laser locked to the OsO$_4$ reference lines listed in Table S1. Those data have been recorded at a trioxane pressure of 1.5 Pa, a pump beam power ranging from $\sim$150~{\textmu}W to $\sim$5 mW (depending on the spectral window probed and as a result of gain-curve-induced power variations). Under these conditions and as illustrated in Fig. \ref{fig:2}, a resolution of $\sim$400 kHz is achieved (2$^{\text{nd}}$-harmonic signal peak-to-peak full width, limited by a combination of collisional, transit-time, and power broadening, as well as FM-induced-distortion). 

\subsubsection{Widely tunable frequency-comb-stabilized-QCL-based spectrometer}
The reader is referred to Refs. \cite{tran2025extending,argence2015quantum,santagata2019high,tran2024near} for a detailed description of the setup and corresponding experimental approaches. In this work, we only briefly present the key aspects. A schematic overview of our QCL spectrometer (Fig. S2) and more details can also be found in the Supplementary Material. The apparatus exploits two SI-traceable and ultrastable frequency references, operated at the LTE (the French time-frequency metrology institute) and transferred to the Laboratoire de Physique des Lasers \textit{via} a 43-km long metrological fibre link of the REFIMEVE infrastructure \cite{cantin2021accurate}. One of these references is an \textit{optical frequency reference} $\nu_{\text{ref}}$ produced by a 1.54 {\textmu}m fiber laser locked to an ultra-stable cavity; the other is a 200 MHz ultrastable \textit{radio frequency (RF) reference} $f_{\text{ref}}$, transferred through the fibre link using an intensity modulated auxiliary 1.54 {\textmu}m laser diode. Both references are calibrated to the primary frequency standards of LTE, consisting in a combination of a liquid-helium cooled cryogenic sapphire oscillator and a hydrogen maser, themselves monitored on a set of atomic fountain clocks.

At LPL, a $\sim$1.54 {\textmu}m laser diode acts as an optical local oscillator (OLO). Using an electro-optic modulator (EOM), a sideband is phase-locked to the remote optical reference. By tuning the microwave frequency driving the EOM, the OLO can be continuously adjusted over a $\sim$10 GHz range. An optical frequency comb (OFC) then transfers the stability, SI-traceability, and tunability of the OLO to a 10.3 {\textmu}m QCL, through a chain of nonlinear conversion and phase-lock loops. The QCL inherits a narrow linewidth ($\sim$0.1 Hz) and a frequency uncertainty below 300~Hz.

Part of the QCL beam is used for saturated absorption spectroscopy of 1,3,5-trioxane in a multi-pass cell. The laser frequency is modulated at 20 kHz (FM amplitude of 400 kHz), and detection is performed at the third harmonic. Spectra of the trioxane $\nu_{5}$ vibrational mode spanning a few gigahertz have been recorded in the regions of the $^{Q}P$(15, $K$), $^{Q}P$(16, $K$), and $^{Q}P$(17, $K$) sub-branches exhibiting $K \leq J$ progressions (see below for transition assignments) at a typical pressure of 1.5 Pa and an intra-cell average power of respectively 0.56 mW, 0.72 mW, and 0.84 mW. Under these conditions and as illustrated in Fig. \ref{fig:3}, a resolution of $\sim$400 kHz is achieved (3$^{\text{rd}}$-harmonic signal peak-to-peak line width, limited by a combination of collisional, transit-time and power broadening, as well as FM-induced-distortion). Previously reported in Refs.\cite{tran2025extending}, these data are further analyzed here.

\section{Mid-infrared spectroscopy of trioxane} \label{sec:3}
\subsection{Survey FTIR spectrum} \label{subsec:3_1}
Fig. \ref{fig:1}(A) displays a survey cell-FTIR spectrum of 1,3,5-trioxane recorded in the 900-1200 cm$^{-1}$ region at 0.1 cm$^{-1}$ resolution. Four strong rovibrational bands are observed, centered around 944, 977, 1071, and 1177 cm$^{-1}$, respectively assigned to the $\nu_{18}$ CH$_{2}$ rocking mode (E symmetry), the $\nu_5$ CO symmetric stretching mode (A$_1$ symmetry) as well as the $\nu_{17}$ and $\nu_{16}$ CO antisymmetric and symmetric stretching modes (E symmetry), in agreement with the assignments from the low resolution gas phase IR spectrum of Kobayashi \textit{et al} \cite{kobayashi1966vibrational}. DFT calculations yield harmonic intensities equal to 64, 82, 71, and 216~km/mol, respectively for these 4 modes, in agreement with our observations. Despite its low resolution, the FT spectrum clearly displays characteristic band contours (parallel band $\nu_{5}$ and perpendicular bands $\nu_{18}$, $\nu_{17}$, and $\nu_{16}$) with partially rotationally resolved P and R branches. Due to the small frequency spacing (about 33~cm$^{-1}$) from $\nu_{18}$ to $\nu_5$ band centers, there is a partial overlap between  $^{P,R}R$ $\nu_{18}$ and $^{Q}P$ $\nu_5$ transitions around 960 cm$^{-1}$. Care will therefore be taken to the potential occurrence of Coriolis coupling between these A$_{1}$- and E-symmetry normal modes in the rovibrational analysis of the $\nu_5$ band in Section \ref{subsec:4_1} .

\subsection{High resolution FTIR spectrum} \label{subsec:3_2}

Fig. \ref{fig:1}(B) displays the cell-FTIR spectrum of 1,3,5-trioxane recorded in the 900-1010~cm$^{-1}$ at 0.001 cm$^{-1}$ resolution. In the 960-1000~cm$^{-1}$ range, it exhibits the intense Q branch of the parallel $\nu_5$ band resulting from stacked $^{Q}Q$ transitions surrounded by $^{Q}P$ and $^{Q}R$ resolved rotational transitions. Fig. \ref{fig:1}(b) also shows the perpendicular $\nu_{18}$ band in the 910-960~cm$^{-1}$ range for which mainly the intense less densely distributed $^{P}P$ and $^{R}R$ transitions are observed and exhibit clear $K \leq J$ progressions. The expanded view in the top of Fig. \ref{fig:1}(b) has been recorded at a lower resolution of 0.02 cm$^{-1}$. It clearly shows a series of hot band Q branches at 975.36, 975.85, 976.20, 976.28, 976.49, and 976.68~cm$^{-1}$, red shifted with respect to the $\nu_{5}$ fundamental Q branch maximum at 977.17 cm$^{-1}$. Their relative intensities are compatible with them being assigned to $n \times \nu_{\text{i}} + \nu_{5} \leftarrow n \times \nu_{\text{i}}$ $(n =1,2)$ transitions starting from the first excited vibrational states of the lowest frequency modes $\nu_{\text{i}}$ such as $\nu_{20}$ (297 cm$^{-1}$), $\nu_7$ (466 cm$^{-1}$) or $\nu_{19}$ (525 cm$^{-1}$), sufficiently populated at room temperature. Although, full assignment of these features is not straightforward, we identify two rather intense hot progressions up to $n=2$ involving the bands at 976.28 and 975.36~cm$^{-1}$ for the first one, those at 976.68 and 976.20 cm$^{-1}$ for the second one, leading to anharmonicities $x_{5i}$ and $x_{5j}$ equal to $-$0.90~(3) and $-$0.49~(1) cm$^{-1}$ between $\nu_5$ and respective low frequency modes $\nu_{i}$ and $\nu_{i}$. The ratios of integrated absorbances between successive bands in these two hot progressions are equal to 0.22 and 0.1 for modes $\nu_{i}$ and $\nu_{j}$ respectively. This agrees well with expected Boltzmann population ratios $e^{\frac{\nu_{i}}{k_{B}T}} \sim0.23$ and  $e^{\frac{\nu_{i}}{k_{B}T}} \sim0.11$ (with $k_B$ the Boltzmann constant and T the temperature) for vibrational modes with frequencies $\nu_{i}$ $\sim$300~cm$^{-1}$ and $\nu_{j}$ $\sim$470 cm$^{-1}$ that could correspond to respectively $\nu_{20}$ and $\nu_7$. A third weak hot band progression is also observed at 976.49 and 975.85 cm$^{-1}$ leading to $x_{5k}$ equal to $-$0.66(3) cm$^{-1}$. Its weakness prevents us from properly assigning the mode involved but $\nu_{19}$ at 525 cm$^{-1}$ is a plausible candidate.

\begin{figure*} \label{fig:1}
\includegraphics{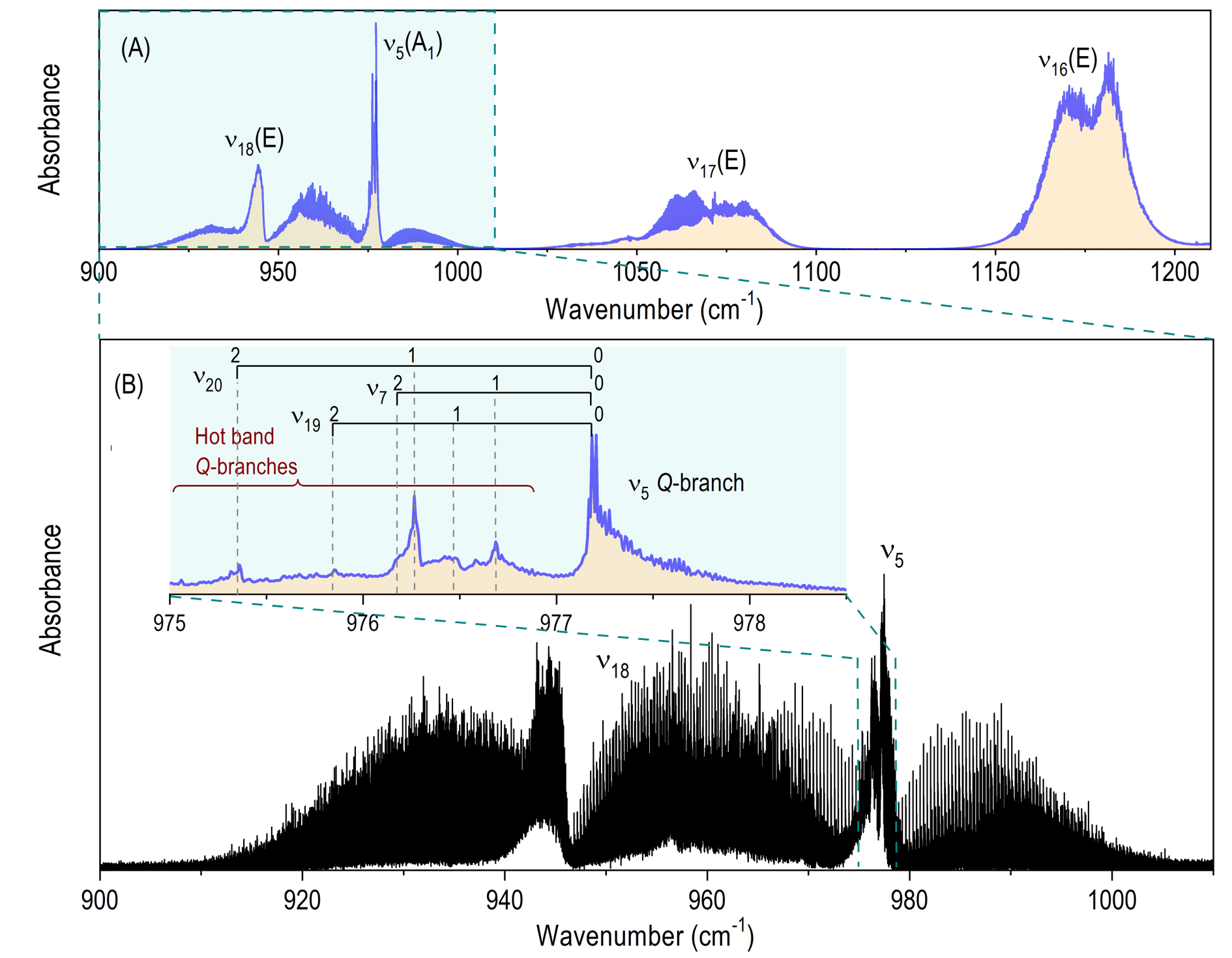}%
\caption{ (a) Room temperature cell-FTIR spectrum of 1,3,5-trioxane recorded at 0.1 cm$^{-1}$ resolution in the 900-1210 cm$^{-1}$ range. (b) Room temperature cell-FT spectrum of 1,3,5-trioxane recorded at 0.001 cm$^{-1}$ resolution in the 900-1010 cm$^{-1}$ range. The top expanded view displays a cell-FT spectrum recorded in the 975-979 cm$^{-1}$ region close to the $\nu_5$ Q-branch at lower resolution (0.02 cm$^{-1}$) to better distinguish the Q-branch of the fundamental band which exhibits an asymmetry with a tail towards the high frequencies (977-979 cm$^{-1}$) and several weaker Q-branches (975-977 cm$^{-1}$), most probably belonging to hot bands.}
\label{fig:1}
\end{figure*}

\subsection{Saturated absorption laser spectra} \label{subsec:3_3}
\begin{figure*} 
\includegraphics[width=0.85\textwidth]{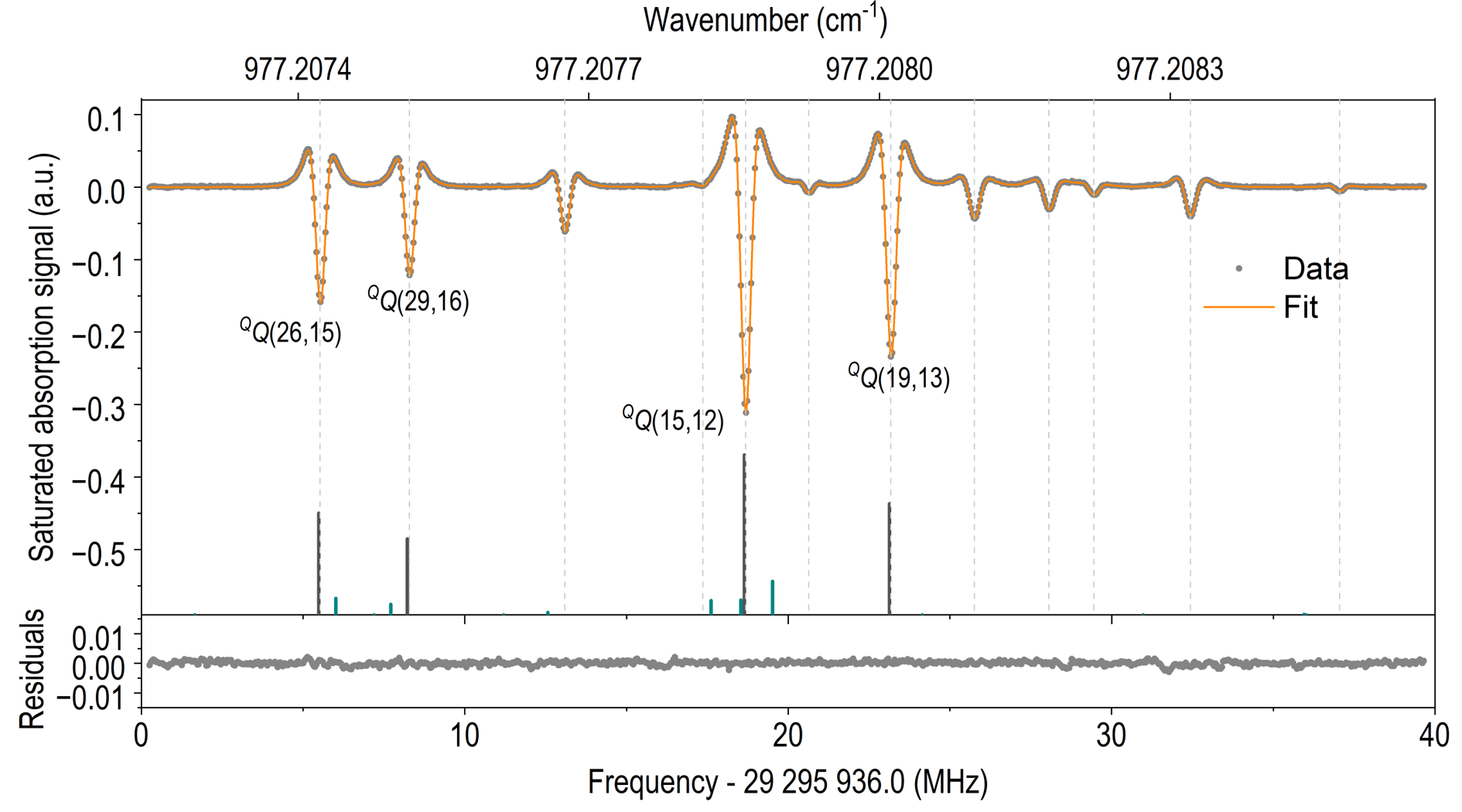}%
\caption{Saturated absorption spectrum of trioxane recorded using the CO$_2$-laser spectrometer (grey dots, average of a pair of opposite frequency sweep direction scans, baseline has been subtracted, see Supplementary Material). The measurement is done in the vicinity of the 10 {\textmu}m $R$(22) CO$_2$ laser emission line at $29 296 129.58537$ MHz. The ultra-stable CO$_2$ laser used is frequency modulated and 2$^{\text{nd}}$-harmonic detection is used. Several isolated lines of the Q branch of the $\nu_5$ vibrational mode are visible. The line assignments $^{Q}Q(J, K)$ resulting from the spectroscopic analysis of Section \ref{subsec:4_2} are labeled. The orange solid line is a fit of 12 line shapes to the spectral contours using a model that takes into account residual amplitude modulation associated with FM and the power frequency-dependency over the scan. Corresponding residuals are shown below. Fitted frequency positions are represented by the dashed vertical lines. Comparison with the calculated stick spectrum using Set 4 of parameters in Table 1 is also shown with the following color code: green sticks correspond to predicted transitions which are not observed in the data and grey sticks to assigned transitions retained in the Hamiltonian fit (not rejected following the 3$\sigma$ criterion, see text). Experimental conditions: pressure, 1.5 Pa; FM frequency, 5015 Hz; FM amplitude, 200 kHz; frequency step, 40 kHz; step duration, 100 ms; lock-in time constant, 100 ms; whole spectrum measurement time, 200 s; pump (resp. probe) beam power of 3 to 4.5 mW (resp. 0.3 to 0.45 mW); beam waist of $\sim$5 mm.}
\label{fig:2}
\end{figure*}

\begin{figure*}
\includegraphics[width=0.85\textwidth]{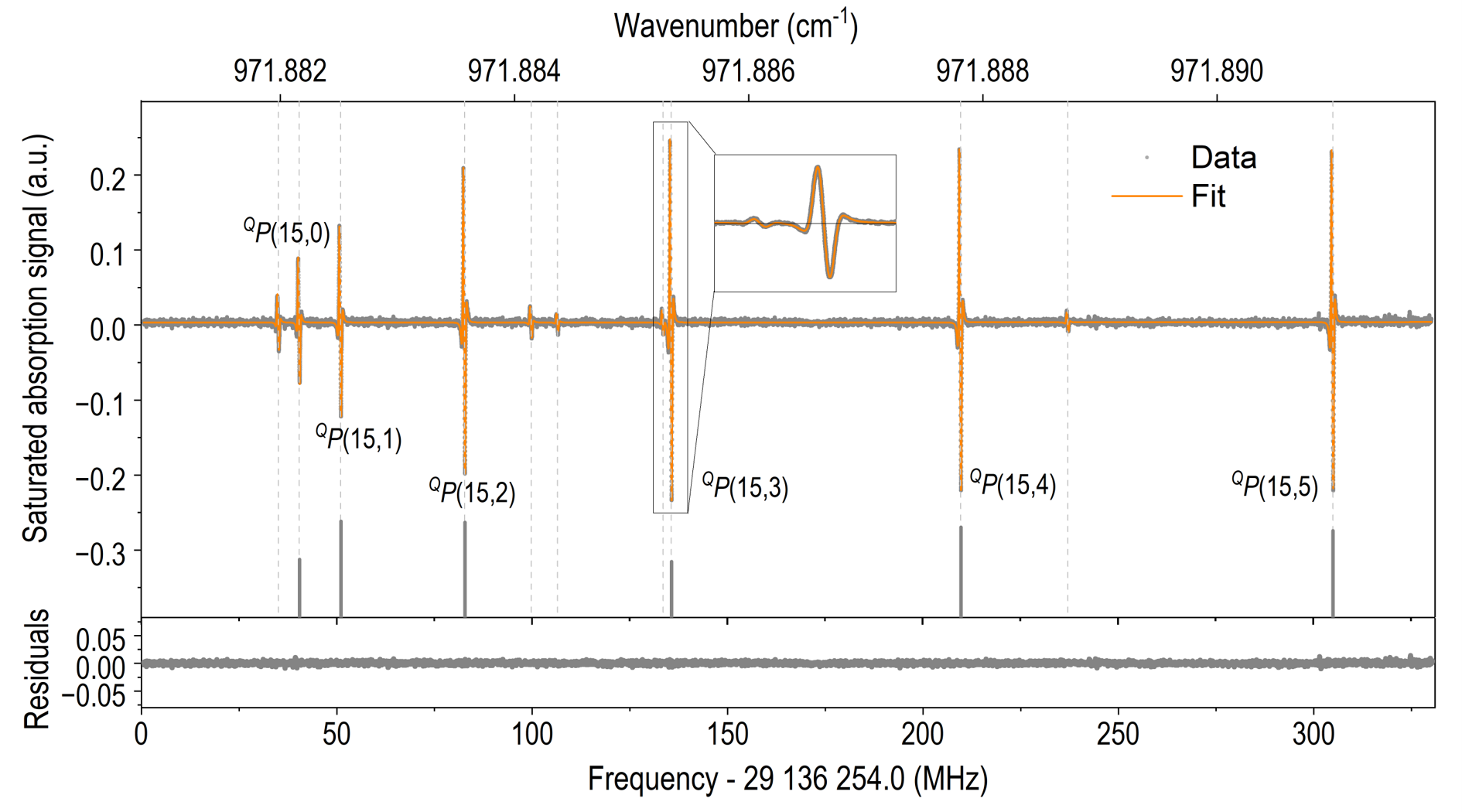}%
\caption{Saturated absorption spectrum of trioxane in the $^{Q}P(15, K)$ $K$-series region ($K$ ranging from 0 to $J$) spanning $\sim$330 MHz recorded with the QCL spectrometer using FM and 3$^{\text{rd}}$-harmonic detection (black dots, average of a pair of up and down scans). The line assignments $^{Q}P(J, K)$ resulting from the spectroscopic analysis of Section \ref{subsec:4_2} are labelled. The orange solid line is a line shape fit of 11 lines to the corresponding spectra using a model that takes into account residual amplitude modulation associated with FM and the power frequency-dependency over the scan. Corresponding residuals are shown below. Experimental frequency positions are represented by dashed vertical lines. The spectrum simulated using Set 4 of parameters in Table \ref{tab:1} is also displayed as grey sticks (all assigned transitions of the spectral window displayed being retained by the corresponding Hamiltonian fit, all sticks are shown in grey, see text). Experimental conditions: pressure, 1.5 Pa; FM frequency, 20 kHz; FM amplitude, 400 kHz; frequency step, $\sim$7.5 kHz; step duration, 5 ms; lock-in time constant, 10 ms; whole spectrum measurement time, 440 s; intra-cell average power of 0.54 mW; beam waist of $\sim$1 mm.}
\label{fig:3}
\end{figure*}

\begin{figure*} 
\includegraphics[width=1\textwidth]{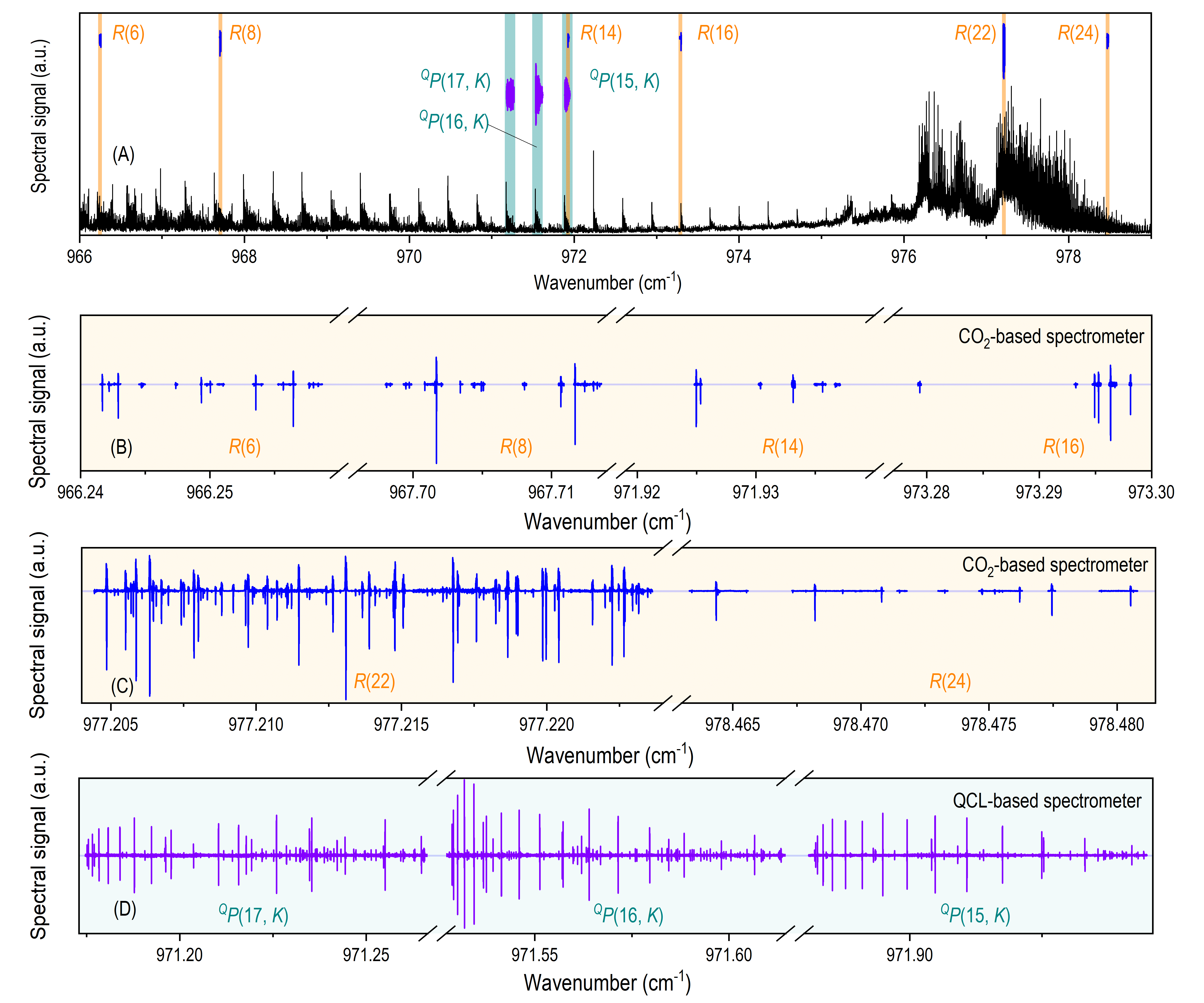}%
\caption{Series of 9 saturated absorption spectra in the $\nu_5$ band of trioxane together with the HR cell-FTIR spectrum. The saturated absorption spectra acquired with the CO$_2$-laser spectrometer are displayed in panels (a), (b), and (c) in blue on an orange background and labeled with the CO$_2$ laser emission lines $R$(6), $R$(8), $R$(14), $R$(16), $R$(22), and $R$(24) in the vicinity of which they have been recorded. They cover the following respective spectral windows spanning between 550 and 600 MHz: 966.2412-966.2595 cm$^{-1}$, 967.6977-967.7164 cm$^{-1}$, 971.9206-971.9399 cm$^{-1}$, and 973.2787-973.2982 cm$^{-1}$ in the  $^{Q}P$ branch (around the $^{Q}P(31,K)$, $(^{Q}P(27,K)$, $^{Q}P(15,K)$ and  $^{Q}P(11,K)$ trioxane $K$-series – $K$ ranging from 0 to $J$ – respectively), and 977.2044-977.2236 cm$^{-1}$ and 978.4628-978.4820 cm$^{-1}$ in the  $^{Q}Q$ branch, allowing to observe both low-$J$ and high-$J$ $^{Q}Q(J,K)$ transitions. The purple data sets on light green background respectively labeled $^{Q}P(17, K)$, $^{Q}P(16, K)$, and $^{Q}P(15, K)$ on panels (a) and (d) have been recorded with the QCL spectrometer. They are labeled with the $^{Q}P(J, K)$ $K$-series in the region of which they have been acquired. They cover the following respective spectral windows spanning between 1.9 and 2.7 GHz: 971.171-971.267 cm$^{-1}$, 971.525-971.615 cm$^{-1}$, and 971.877-971.944 cm$^{-1}$.}
\label{fig:5}
\end{figure*}

\begin{figure*} 
\includegraphics[width=0.9\textwidth]{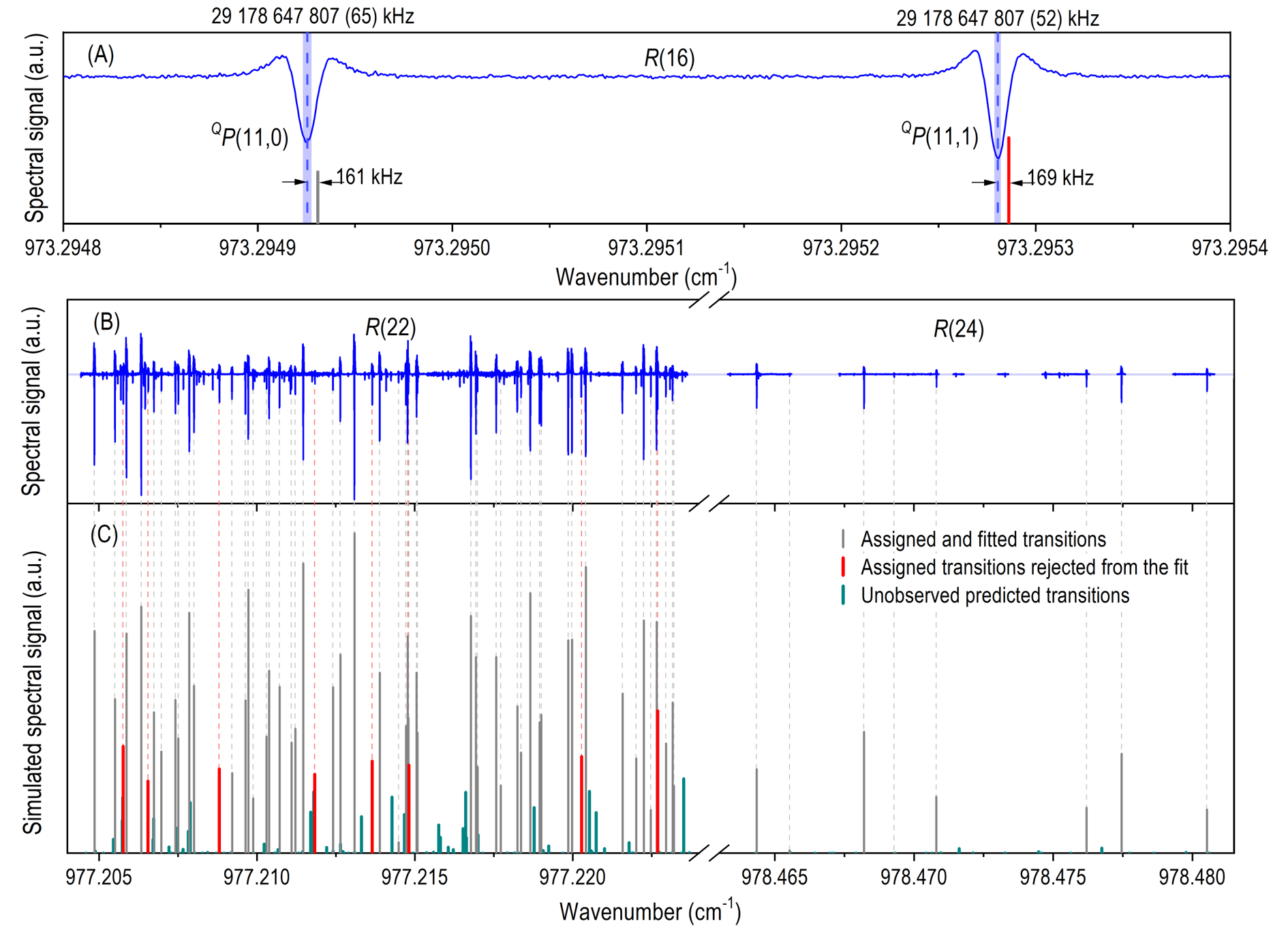}%
\caption{Comparison between the calculated stick spectrum using Set 4 of parameters in Table 1 and saturated absorption spectra recorded around the $R$(16), $R$(22), and $R$(24) CO$_2$ laser emission lines (in panel (a), (b), and (c) respectively, data in blue already shown in Fig. \ref{fig:5} and partially in Fig. \ref{fig:2}). The color code for the simulated stick spectrum is the following: green corresponds to predicted transitions which are not observed in the data, red to transitions assigned in the experimental spectra but rejected from the Hamiltonian fit following the 3$\sigma$ criterion (see text) and grey to assigned transitions retained in the fit. Dotted vertical lines pin point assigned transitions with the same color code. In panel (a) which focuses on two isolated transitions, assignments are shown. Fitted frequency positions and corresponding uncertainties are given at the top and represented by the dashed vertical lines and the surrounding light blue rectangle.}
\label{fig:6}
\end{figure*}

Fig. \ref{fig:2} shows a spectrum spanning 40 MHz in the vicinity of the $R$(22) CO$_{2}$ laser emission line recorded using the CO$_2$-laser spectrometer and revealing the distinctive rich structure of relatively well isolated lines of the $\nu_5$ vibrational mode $Q$ branch. Saturated absorption spectra spanning $\sim$600 MHz have been recorded around each of the CO$2$ laser emission lines listed in Section \ref{subsec:2_1}, leading to a total span of 3.6 GHz in a $\sim$400-GHz-wide region. Fig. \ref{fig:3} shows a $\sim$330-MHz-wide spectrum of the $^{Q}P(15, K)$ sub-branch (see Section \ref{sec:4} below for the details of the assignments) obtained with the QCL spectrometer. We have accumulated such spectra spanning a total of 7.5 GHz, in a region covering $\sim$25 GHz. All the saturated absorption laser spectra exploited in this work are shown in Fig. \ref{fig:5}. Fig. \ref{fig:6} also displays a couple of measurements. The data recorded using the QCL spectrometer displayed in Fig. $\ref{fig:3}$, and in purple in Fig. \ref{fig:5} have also already been reported in Ref. \cite{tran2025extending}.

\begin{figure*} 
\includegraphics[width=0.9\textwidth]{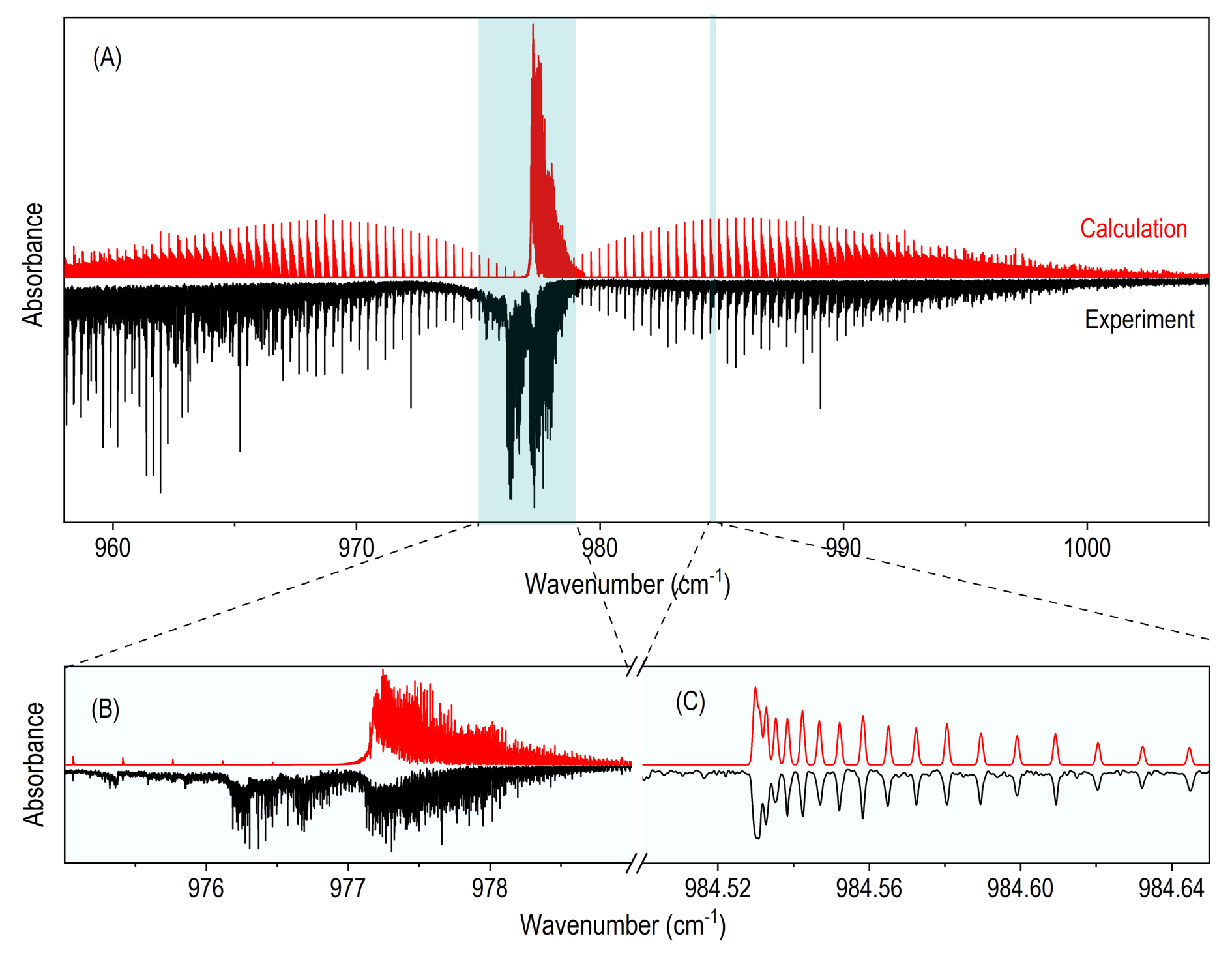}%
\caption{(a) Room temperature cell-FTIR spectrum of the $\nu_5$ band of trioxane (partially overlapping with the $\nu_{18}$ band in the 958-963 cm$^{-1}$ range) compared to the simulation using Set 2 of molecular parameters in Table \ref{tab:1} resulting from the combined Hamiltonian fit of FTIR (this work) and mmw \cite{klein1996terahertz} data.  The $\nu_5$ band spectrum simulated at 300 K is convoluted with a Gaussian line shape of 0.0012 cm$^{-1}$ width. (b) The expanded view of the  $^{Q}Q(J,K)$ fundamental transitions shows the good agreement between observed and calculated rovibrational band contours with an asymmetry to the red side, while, in the experimental spectrum, several hot bands blur the $^{Q}P$ transitions at low $J,K$ values. (c) The expanded view of the $^{Q}R(20,K)$ $K$-series transitions (top right) illustrates the slight intensity differences resulting from spin statistics with $K = 0$, $K = 3n$, and $K \neq 3n$ transitions ($K$ being the ground state rotational quantum number and \textit{n} a non-zero integer) having weights of respectively 24, 48, and 40.}
\label{fig:4}
\end{figure*}

To compensate for frequency shifts caused by our finite detection bandwidth \cite{rohart2014absorption,rohart2017overcoming}, the QCL frequency is swept in both directions with increasing and decreasing frequency, and each spectrum consists in the average of such a pair of up and down scans. To increase signal-to-noise ratio, the FM amplitude has been set to a few 100 kHz. But at this level, observed line shapes are distorted away from the typical Lorentzian profile, as seen in Fig. \ref{fig:2}. Note in particular that they become asymmetrical, making it harder to pinpoint the line centre. Several effects can contribute to this, the likeliest being residual amplitude modulation associated with FM and the power variation over a scan (resulting from the laser gain curve, the non-trivial Doppler contour or residual interference fringes). We thus fit to our spectra a theoretical line shape model for FM spectroscopy inspired by Refs. \cite{arndt1965analytical,schilt2003wavelength}, and already exposed in Ref. \cite{tran2024near} that takes the combined occurrence of intensity and frequency modulation and the associated distortions into account, after demodulation on any given $n^{\text{th}}$ harmonic. Fitting to the spectral contours our 2$^{\text{nd}}$- or 3$^{\text{rd}}$-harmonic detection line shape model allows us to determine the line-centre frequency of each transition with an uncertainty ranging from 40 kHz to 200 kHz for assigned transitions (see section \ref{sec:4} below for the details of the assignments). Fig. \ref{fig:2}  and \ref{fig:3} show the resulting line shape fit and corresponding residuals, together with the fitted frequency positions shown as dashed vertical lines (see also Fig. \ref{fig:6}(a) in which experimental frequency uncertainties are also shown as light blue rectangle). Details on the data analysis, on the procedure followed for fitting spectra and on the uncertainty budget are given in the Supplementary Material.

A total of 557 saturated absorption lines have been recorded. Tables S3 of the Supplementary Material lists the line-centre absolute frequencies and uncertainties of the 257 transitions measured with the CO$_2$-laser spectrometer. As detailed in our previous work \cite{tran2025extending}, amongst a total of 312 lines observed using the QCL spectrometer, 47 were assigned (see section \ref{sec:4} below for assignments). The corresponding list of assignments, frequencies and uncertainties can be found in Table S4 (details on the 265 unassigned other lines are given in Ref. \cite{tran2025extending}). Finally, Tables S5 lists the line-centre absolute frequencies and global uncertainties of the 12 rovibrational transitions which have been recorded with both spectrometers.

\section{Rovibrational analysis of the $\nu_5$ CO symmetric stretching mode of trioxane}\label{sec:4}

The rovibrational spectrum of the $\nu_5$ parallel band of trioxane is analyzed with the oblate symmetric rovibrator model, including the rotational constants $B$ and $C$, the quartic ($D_J$, $D_{JK}$, and $D_K$) and sextic ($H_J$, $H_{JK}$, $H_{KJ}$, and $H_K$) centrifugal distortion constants. Ground state rotational energy levels are given by:

\begin{widetext}
\begin{equation}
\label{eq:1}
\begin{split}
E_{\text{gr}}(J, K) = &BJ(J + 1) + (C - B)K^2 - D_J J^2(J + 1)^2 - D_{JK}J(J + 1)K^2 - D_K K^4 \\
&+ H_J J^3(J + 1)^3 + H_{JK}J^2(J + 1)^2 K^2 + H_{KJ}J(J + 1)K^4 + H_K K^6
\end{split}
\end{equation}
\end{widetext}

The rovibrational energy levels in the first excited vibrational state of the $\nu_5$ mode are calculated from the following expression (in which $\nu_5$ is the band center and molecular parameters are primed):

\begin{widetext}
    \begin{equation}
\begin{split}
E_{\nu_5}(J, K) = \nu_5 &+ B'J(J + 1) + (C' - B')K^2 - D'_J J^2(J + 1)^2 - D'_{JK}J(J + 1)K^2 - D'_K K^4 \\
&+ H'_J J^3(J + 1)^3 + H'_{JK} J^2(J + 1)^2 K^2 + H'_{KJ}J(J + 1)K^4 + H'_K K^6
\end{split}
\label{eq:2}
\end{equation}
\end{widetext}

Transitions for such a parallel band follow the selection rules $\Delta J=0,\pm1$ and $\Delta K=0$, and are designated by $^{Q}P(J,K)$, $^{Q}Q(J,K)$, and $^{Q}R(J,K)$ where the superscript Q corresponds to $\Delta K=0$, the main capital letter $P$, $Q$, $R$ to $\Delta J = -1, 0, 1$, respectively, and where $J$ and $K$ are the ground state rotational quantum numbers. The lower-state statistical weights are affected by the $2J + 1$ rotational degeneracy factor and the rules of nuclear spin statistics for a $C_{3v}$ molecule possessing six identical H atoms of nuclear spin $I_H = 1/2$ outside the axis of symmetry. As a result, all transitions with $K = 3n$ ($n$ is an integer different from zero) carry the nuclear spin weight 48 (24 for $K = 0$), whereas all transitions with $K \neq 3n$ have the weight 40.9.

In the following rovibrational analyses of the $\nu_5$ CO symmetric stretching mode, we want to combine our mid-IR data to ground state measurements from the literature. A ground state rotational spectroscopic study of trioxane in the mmw range has been reported by Klein \textit{et al}  in the 326-947 GHz range: rotational transitions with $J$ values ranging from 30 to 89 and $K$ values up to 77 were measured at an accuracy better than 10 kHz. The corresponding ground state rotational constants ($B$, $D_J$, $D_{JK}$, $H_J$, $H_{JK}$, and $H_{KJ}$) are designated as Set 1 and given in Table \ref{tab:1}.

For parallel bands of symmetric top molecules, the excited state constants $C^\prime$, $D_K^\prime$ and $H_K^\prime$ parametrize contributions to the Hamiltonian which are only $K$-dependent and so cannot be determined independently from their ground state counterpart. Consequently, only the differences $\Delta C=C^\prime-C$, $\Delta D_K = D_K^\prime - D_K$, and $\Delta H_K=H_K^\prime-H_K$ between the excited and ground states will be determined in the present work.

In the following, we analyze the rovibrational spectrum of the $\nu_5$ symmetric CO stretching mode by progressively refining the rovibrational Hamiltonian introduced above against the measured transition frequencies. The determination of reliable molecular parameters requires an unambiguous assignment of the observed  spectral lines to the transitions predicted by the model. To achieve this, the spectroscopic data are treated in three successive steps. First, we analyze our FTIR data together with Klein \textit{et al}’s mmw \cite{klein1996terahertz} data. This provides reliable rovibrational assignments, and an initial determination of the excited-state spectroscopic constants. Next, we analyze our saturated absorption lines with the mmw data. Although this involves far fewer rovibrational transitions, its orders-of-magnitude higher frequency accuracy, enables a substantial refinement of the excited-state parameters. Finally, all three data sets are combined in a global Hamiltonian fit. Besides yielding the most accurate set of molecular constants, this global analysis allows us to determine the correction to apply to the FTIR frequency-scale by identifying the calibration that maximizes the agreement of the complete spectroscopic data set with the Hamiltonian model. The corrected FTIR transition frequencies are then incorporated into the final fit used to derive the molecular constants. All analyses are performed using the PGOPHER program \cite{western2017pgopher} and the SPFIT/SPCAT \cite{pickett1991fitting} suite throughout. During the fitting procedure, assigned lines may be excluded according to a standard $3\sigma$ rejection criterion whereby any transition for which the difference between the experimental and calculated frequency exceeds three times the experimental precision ($\sigma$) is discarded. Line lists of the assigned transitions are provided in the Supplementary Material.

\subsection{Rovibrational analysis of the cell-FTIR spectrum}\label{subsec:4_1}

The FTIR rovibrational analysis of the $\nu_5$ band of 1,3,5-trioxane was facilitated by the series of $^{Q}P$ and $^{Q}R$ rotational transitions exhibiting clear-cut resolved K sub-structures in the \textit{P}- and \textit{R}-branches with maximal values of $J$ reaching 64 in the \textit{R}-branch and 43 in the $P$-branch. So, in a first step, a straightforward assignment could be realized for both branches. Note that a local Coriolis-type vibrational perturbation was evidenced in the range 991-996 cm$^{-1}$ for the $^{Q}R(J,K)$ transitions with  $J=43-56$ and $K\geq 25$ leading to frequency deviations reaching 0.015 cm$^{-1}$ with respect to the rigid rotor model. Corresponding perturbed lines were discarded from the simulation. The $(\nu_7 + \nu_{19})$ combination band of $E$ symmetry expected around 992 cm$^{-1}$ is most probably the perturbing state but the treatment of this perturbation is beyond the scope of the paper. Conversely, no deviation was observed for the $^{Q}P$ transitions with $J \geq 40$ in overlap, as pointed out in Section \ref{subsec:3_1}, with the $^{P,R}R$ transitions of the $\nu_{18}$ band around 960 cm$^{-1}$. It happens thus not necessary to introduce any Coriolis interaction between $\nu_5$ and $\nu_{18}$ bands.

Following a preliminary Hamiltonian fit, we were able to complete the list of assigned lines with $^{Q}Q$ transitions of $J$ and $K$ values up to 85, which enabled us to adjust centrifugal distortion parameters up to the sextic order. Fig \ref{fig:4} displays the comparison between the experimental FT mid-IR spectrum of the $\nu_5$ symmetric band and the simulation of its contour (fundamental band alone). Set 2 in Table \ref{tab:1} gathers the ground and $\nu_5=1$ state parameters derived from the global fit of 2142 FTIR and 218 mmw assigned lines. It yields a root mean square (RMS) of differences between observed and calculated frequencies (obs.-calc.) of 10 kHz for mmw lines and 0.00018 cm$^{-1}$ for mid-IR lines, after application of the $3\sigma$ rejection criterion ($\sigma$ is generally fixed to 0.0002 cm$^{-1}$ for cell-FTIR experiments at 0.001 cm$^{-1}$ resolution, see also line lists in the Supplementary Material). Here 11 mmw and 20 mid-IR lines have been rejected (giving a total of 207 mmw and 2122 mid-IR retained lines, respectively).

\begin{table*}
\centering
\caption{Ground- and excited-state parameters derived from four different sets of measured transitions in the ground vibrational state and the $\nu_5$ band of trioxane. Set 1 is derived from the mmw study of Klein \textit{et al} \cite{klein1996terahertz}, Set 2 from the combination of Klein \textit{et al} mmw study and the cell–FTIR spectrum, Set 3 from the combination of Klein \textit{et al} mmw study and the sub-Doppler measurements, and Set 4 from the combination of Klein \textit{et al} mmw study and all FTIR and sub-Doppler lines. The uncertainties in parentheses correspond to one $\sigma$ standard deviations. Table S6 of the Supplementary Material shows the same information as Table \ref{tab:1}, but presented in a way that allows a quick comparison of how a single parameter and its uncertainty change from set to set (see also Fig. S5 to S6).}
\begin{ruledtabular}
\renewcommand{\arraystretch}{1.1}
\begin{tabular}{lcccc}
\toprule
 & Set 1 & Set 2 & Set 3 & Set 4 \\
 \hline
\midrule
$B/\text{MHz}$ & 5 273.257 180 (33) & 5 273.257 183 (37) & 5 273.257 166 (47) & 5 273.257 171 (40) \\
$D_J/\text{kHz}$ & 1.343 879 7 (80) & 1.343 890 5 (83) & 1.343 886 5 (105) & 1.343 888 3 (90) \\
$D_{JK}/\text{kHz}$ & $-$2.016 295 (17) & $-$2.016 340 (19) & $-$2.016 338 (23) & $-$2.016 340 5 (197) \\
$H_J/\text{mHz}$ & 0.490 61 (55) & 0.491 41 (55) & 0.491 14 (69) & 0.491 29 (59) \\
$H_{JK}/\text{mHz}$ & $-$2.097 8 (15) & $-$2.099 2 (13) & $-$2.098 8 (16) & $-$2.099 2 (14) \\
$H_{KJ}/\text{mHz}$ & 2.740 8 (34) & 2.735 7 (18) & 2.735 2 (22) & 2.735 6 (19) \\
$\nu_5/\text{cm}^{-1}$ & & 977.171 151 (13) & 977.170 979 14 (42) & 977.170 979 81 (36) \\
$B'/\text{MHz}$ & & 5 271.464 82 (96) & 5 271.469 328 (74) & 5 271.469 397 (61) \\
$\Delta C/\text{MHz}$ & & 8.944 51 (88) & 8.969 797 (93) & 8.969 528 (70) \\
$D_J'/\text{kHz}$ & & 1.33187(65) & 1.332 442 (87) & 1.332 800 (50) \\
$D_{JK}'/\text{kHz}$ & & $-$1.469 3 (15) & $-$1.438 62 (38) & $-$1.44 042 (17) \\
$\Delta D_K/\text{kHz}$ & & $-$0.209 61 (98) & $-$0.235 86 (41) & $-$0.234 055 (169) \\
$H_J'/\text{mHz}$ & & 0.16 (12) & 0.251 (14) & 0.221 6 (92) \\
$H_{JK}'/\text{mHz}$ & & $-$4.85 (40) & $-$0.434 (78) & $-$0.126 (42) \\
$H_{KJ}'/\text{mHz}$ & & 32.10 (56) & 26.74 (18) & 25.937 (79) \\
$\Delta H_K/\text{mHz}$ & & $-$17.565 (263) & $-$16.48 (14) & $-$15.880 (57) \\
\bottomrule
\end{tabular}
\end{ruledtabular}

\label{tab:1}
\end{table*}

The asymmetry of the $Q$ branch observed in Fig. \ref{fig:1}(b) can be linked to the sign of the rovibrational energy term $asymQ = J(J+1)\Delta B + K^2 (\Delta C - \Delta B)$ where $\Delta B= B^{\prime}-B$. According to Set 2 in Table \ref{tab:1}, $asymQ \sim 10.8 K^{2} - 1.8 J^{2}$. A careful examination of the $^{Q}Q$ branch shows that the most intense lines satisfy $K > 0.5 J$ and thus $asymQ > 0$, confirming the asymmetry of the $Q$ branch resulting in a tail towards the high frequencies.

\subsection{Rovibrational analysis of all saturated absorption data} \label{subsec:4_2}

Fig. \ref{fig:5} displays the series of saturated absorption spectra recorded with both the CO$_2$-laser and QCL spectrometers in the 966-979 cm$^{-1}$ region of the $^{Q}P$ and $^{Q}Q$ branches of the $\nu_5$ band together with the FTIR spectrum. Frequency positions obtained using the parameters of Set 2 derived from the preceding FTIR analysis were sufficiently predictive to assign 123 saturated absorption lines (67 $^{Q}Q(J,K)$ and 56 $^{Q}P(J,K)$ lines, cf Table S3 and S4 in the Supplementary Material) out of the 557. This step highlights the importance of the FTIR spectrum as an essential precursor for sub-MHz spectroscopic characterization. Two lines were recorded with both spectrometers, for which the QCL measurements, exhibiting the lower uncertainty, have been used in subsequent analyses (cf Table S5 in the Supplementary Material).

A global Hamiltonian fit to the mmw and saturated absorption data including 208 mmw of Klein \textit{et al} \cite{klein1996terahertz} and 112 sub-Doppler mid-IR lines (after rejection of 10 mmw and 11 saturation lines by the $3\sigma$ criterion) provided improved determinations of the ten excited state parameters of the $\nu_{5} = 1$ state with a normalized RMS of (obs.-calc.) frequencies equal to 1.14. The resulting parameter set is reported in Table 1 as Set 3. As shown in Fig. \ref{fig:2} and \ref{fig:3}, unassigned lines are observed. The $\sim$400 mostly weak unassigned lines most probably belong to hot bands or to fundamental bands of less abundant isotopologues, the study of which is beyond the scope of the present paper.

\subsection{Combined FTIR and saturated absorption data analysis} \label{subsec:4_3}

To further improve the precision of excited state parameters, the mmw line positions from Klein \textit{et al} \cite{klein1996terahertz} are combined with infrared line positions derived from both FTIR and saturated absorption spectra in a global Hamiltonian fit of the $\nu_5$ band of trioxane.

As shown in Table \ref{tab:1} (and Table S6, and Fig. S6 of the Supplementary Material), parameters extracted from analyzing mid-IR FTIR and sub-Doppler data sets (Sets 2 and 3 respectively) do not seem to agree within their error bars $\sigma$. They can in fact differ by many $\sigma$, as exemplified by the $1.7 \times 10^{-4}$ cm$^{-1}$ difference for $\nu_5$, corresponding to almost $\sim13\sigma$. This level of discrepancy can be caused by frequency shifts in the original FTIR spectra, even small ones that are within the calibration error of the instrument (Section \ref{subsec:2_1}). This highlights the usefulness of metrology-grade sub-MHz laser spectroscopy in ascertaining molecular parameters. We note here that all our laser spectroscopy data recorded using both laser spectrometers agree to within their error bars, and differ only by a few tens of kilohertz (see Table S5 and accompanying text in the Supplementary Material).

We therefore use the sub-Doppler measurements to calibrate the FTIR data a posteriori. The FTIR frequency-scale correction is determined within a global Hamiltonian analysis by exploiting the complete spectroscopic data set. Since residual calibration errors in Fourier-transform spectroscopy are expected to arise predominantly from a multiplicative error in the frequency scale, the correction is assumed to be proportional to the transition frequency. Accordingly, 121 corrected replicas of the FTIR line list are generated by applying multiplicative frequency-scale corrections to the original transition frequencies $ \tilde{\nu}_{\text{FTIR}}^{0}$ according to:
\begin{equation}
  \tilde{\nu}_{\text{FTIR}}^{\delta} = \tilde{\nu}_{\text{FTIR}}^{0} \left( 1 + \frac{\delta}{ \tilde{\nu}_{\text{c}}} \right)  
\end{equation}

where $\tilde{\nu}_c$  $\sim$981 cm$^{-1}$ is chosen close to the center of the $\nu_5$ band (962-1000 cm$^{-1}$). The trial corrections correspond to frequency shifts $\delta$ ranging from $-8 \times 10^{-4}$ to $+4 \times 10^{-4}$ cm$^{-1}$, by steps of $10^{-5}$~cm$^{-1}$ ($\sim$0.3 MHz). Each corrected FTIR line list is then combined with the millimetre-wave and sub-Doppler data in an independent global Hamiltonian fit. The optimal frequency-scale correction is identified by comparing the 121 fits according to several complementary statistical criteria that are minimized simultaneously: (i) the normalized RMS of (obs.-calc.) frequencies; (ii) the number of rejected lines with particular attention to the high $J$ and $K$, $^{Q}Q(J,K)$ sub-Doppler transitions recorded near the $R$(24) CO$_{2}$ laser emission line, which provide strong constraints on the sextic centrifugal distortion constants; and (iii) the uncertainties of the fitted excited-state parameters.

This procedure leads to an optimal frequency-scale correction corresponding to a shift of $\delta=(-5.0 \pm 0.5) \times 10^{-5}$ cm$^{-1}$ applied to the FTIR line list, where the quoted uncertainty is deliberately conservative (see the Supplementary Material). The resulting global fit yields a normalized (obs.-calc.) RMS of 0.983 (6.3 MHz for the mid-IR lines), with 103 rejected lines, including 79 FTIR, 11 mmw, and 13 sub-Doppler lines. The optimal correction and its uncertainty are approximately 4 and 40 times smaller, respectively, than the FTIR spectrometer calibration accuracy discussed in Section II-1. Compared with this conventional calibration procedure, the present \textit{a posteriori} approach offers two major advantages. First, it exploits the full metrological accuracy of approximately $\sim100$ SI-traceable sub-Doppler transitions. Second, the global Hamiltonian analysis propagates these absolute frequency references throughout the entire FTIR broadband data set, allowing more than two thousand FTIR transitions to contribute simultaneously to the determination of the frequency-scale correction. Consequently, the correction is constrained by about fifty times more experimental information than in the conventional calibration, which relies on only 39 FTIR transitions, thereby substantially improving the calibration sensitivity. This strategy particularly benefits from the high spectral density of a relatively large polyatomic molecule such as trioxane, whose room-temperature spectrum contains thousands of rotationally resolved rovibrational transitions. Additional details on the calibration procedure are provided in the Supplementary Material. The corresponding line list and differences between the observed and calculated transition frequencies are available in Table S7 of the Supplementary Material. The final global Hamiltonian fit includes, after rejection of lines by the $3\sigma$ criterion, 2380 lines (207 mmw, 2063 FTIR and 110 sub-Doppler lines). The corresponding molecular parameters are gathered in Set 4 of Table \ref{tab:1}. Fig. \ref{fig:2}, \ref{fig:3}, and \ref{fig:6} show the very good agreement between the saturated absorption experimental data and the simulated spectrum in the regions of the $R$(16), $R$(22), and $R$(24) CO$_2$ laser emission lines and in the $^{Q}P(15, K)$ sub-branch of trioxane. Retained and rejected assigned transitions are respectively shown as grey and red vertical sticks. Green vertical sticks correspond to unobserved predicted transitions.

When performing this global fit including mmw data of Klein \textit{et al} \cite{klein1996terahertz}, all molecular parameters (except $C$ and $D_K$ in the ground state) were free to be adjusted, enabling us to slightly refine the ground-state constant respect to the pure mmw analysis. The ground state parameters obtained in combined and pure mmw analyses coincide well within their error bars. Another ground state rotational mmw spectroscopic study has been reported: compared to Klein \textit{et al} \cite{klein1996terahertz}, Gadhi \textit{et al} \cite{gadhi1989submillimeter} probed lower values of $J$ ranging from 0 to 47 and $K$ values up to 37 in the 10-505 GHz range at a 10 times lower accuracy (<100 kHz). Surprisingly, those two studies do not agree as evidenced by ground state rotational parameters which do not coincide within their error bars (see the example of $H_{JK}$ in Fig. S5 of the Supplementary Material). It is a priori difficult to decide which of the two is more reliable. However, conducting global fits after replacing Klein \textit{et al} mmw data by those of Gadhi et al, not only results in $\sim$10 times degraded ground state rotational constant uncertainties, but also leads to several ground state parameters ($D_{JK}$, $H_{JK}$, and $H_{KJ}$) which do not coincide well, within their error bars, with those obtained from Gadhi \textit{et al}’s pure mmw study. Such inconsistencies are much less pronounced when using Klein \textit{et al}’s data. Fig. S5 in the Supplementary Material illustrates this by comparing the values of $H_{JK}$ obtained following different Hamiltonian fits using either set of mmw data. This allowed us to discriminate between the two mmw studies and prompted us to consider Klein \textit{et al}’s line list as more reliable. 

\section{Discussion} \label{sec:5}

According to the above, only $\sim$110 mid-IR line positions determined with record few 10 kHz uncertainties using sub-Doppler laser spectroscopy have enabled us to determine molecular parameters with 3 to 30 times improved uncertainties compared to what is achieved with $\sim$2500 frequencies known to within 6 MHz by FT spectroscopy. However, our work demonstrates that the two data sets are required to obtain the most accurate set of molecular parameters. Saturated absorption data is essential not only to achieve the highest accuracies, but also more pragmatically for conducting the \textit{a posteriori} frequency calibration of FT data which is an integral part of the procedure we establish and without which the combined analysis would not be possible. Conversely, frequency positions obtained using the parameters of Set 2 derived from the FTIR analysis are also crucial for primarily unlocking the assignment of saturated absorption spectra and thus for their exploitation. Furthermore, if parameters Set 2 and Set 3 are largely incompatible as a result of the calibration imperfections of the FT spectrometer, even after post-calibration, parameters in Set 3 and Set 4, although much more consistent, are not fully compatible within their error bars. The spectral coverage of our ultra-precise sub-Doppler measurements has only allowed transitions in the $P$ and $Q$ branches to be probed which we believe leads to biases in the determination of molecular parameters. In fact, our previous sub-Doppler study of $\nu_5$ fundamental band, which relied only on the 47 assigned $P$-branch lines from the QCL spectrometer data (see Table S4 of the Supplementary Material), yielded even more biased parameters (in addition to uncertainties one to two orders of magnitude larger) \cite{tran2025extending}. Complementing our full metrology-grade data set with a HR FT study providing thousands of lines appropriately sampling the entire $P$, $Q$, and $R$ vibrational landscape from low to high $K$ and $J$ is key to avoid biases. This explains the slight inconsistency between Set 3 and Set 4. Both sets display similar uncertainties – Set 4 has slightly better uncertainties for quartic and sextic constants – but the latter exhibits more accurate excited state molecular parameters and reduced systematic shifts. The comparison between the different sets of parameters exposed in this paragraph is facilitated by Table S6, and Fig. S5 and S6 in the Supplementary Material.

This has allowed us to improve our understanding of the rovibrational structure of trioxane, to bring new spectroscopic information and build a model supported by molecular parameters with record uncertainties for a molecule of this size: a relative error of $3.7\times 10^{-10}$ for the band center, $\sim 10^{-8}$ for rotational constants in the $\nu=1$ vibrationally excited state, from 0.004 to 0.012$\%$ for the quartic constants ($D_J^{\prime}$, $D_{JK}^{\prime}$, $\Delta D_K$) and, for the sextic parameters, from 0.3$\%$ ($H_{KJ}^{\prime}$, $\Delta H_K$) to 4.2$\%$ ($H_J^{\prime}$) and 33$\%$ ($H_{JK}^{\prime}$).

This has been made possible because (i) we have demonstrated mid-IR sub-Doppler spectroscopy for, to our knowledge, the largest molecule to date leading to record uncertainties for such a complex molecule of a few 10 kHz, $\sim10^{-9}$ in relative value, depending on the line intensity, and (ii) we have established procedure, based on a metrology strategy for accurate frequency calibration, allowing us to extract very accurate molecular parameters of a large polyatomic species, both in the ground and excited states, from a large amount of spectroscopic data spanning a wide spectral window, taken at different resolutions on complementary experiments. This highlights the potential of extending frequency metrology to increasingly larger polyatomic molecules in combination to more conventional approaches.

In addition to providing models of unprecedented accuracies for increasingly complex species, it opens perspectives for exploring subtle intramolecular coupling mechanisms. Measured line shapes are fully compatible with the expected combination of pressure, transit-time and power broadening, as well as frequency-modulation-induced distortions. Our data thus do not show any signature of strong intramolecular vibrational redistribution (IVR) \cite{nesbitt1996vibrational} or other subtle rovibrational coupling mechanisms \cite{liu2023ergodicity} at play that would manifest as additional distortions or the presence of multiplets (see also the discussion in our previous work \cite{tran2025extending}). This is not necessarily a surprise for a rigid and symmetric molecule such as trioxane probed at 10 µm corresponding to a relatively low internal energy and with only three low-lying vibrational modes below 600 cm$^{-1}$ of which the lowest one falls around 300~cm$^{-1}$. However, although we consider our frequency metrology and SI-traceable approach to be highly rigorous and our sub-Doppler data assignments to be unequivocal, a few lines, as shown in red in Fig. \ref{fig:6}(a), are surprisingly rejected by the Hamiltonian fit after application of the $3\sigma$ rejection criterion. As illustrated in Fig. \ref{fig:6}(a) (see also Fig. \ref{fig:2} and \ref{fig:3}), the density of lines is sufficiently low for the coincidence between modelled and measured lines to be unambiguous, with no other surrounding transitions that could make it disputable. However, the observed transition assigned to $^{Q}P(11,1)$ (experimental precision $\sigma$ = 52 kHz, represented by the light blue rectangle around the dashed line corresponding to the fitted frequency position) is rejected from the Hamiltonian fit with a difference (obs.-calc.) of 169 kHz larger than 3$\sigma$. Transition $^{Q}P(11,0)$, on the other hand, successfully meets the 3$\sigma$ criterion and is retained. In addition, as shown in green Fig. \ref{fig:2}, Fig. \ref{fig:6}(b) and (c), several lines of the $Q$ branch predicted in the model could not be reasonably assigned. Most of these are low intensity transitions and may be buried in the noise. A few with a priori measurable intensities did not coincide with measured resonances, those being too far. We believe the above discrepancies could be signatures of the onset of perturbations induced by intramolecular couplings (Coriolis coupling, Fermi resonances…) that the Hamiltonian fitted to the data does not take into account. These are probed here at an unrivalled level of resolution, corresponding to shifts smaller than $\sim$100 kHz. This shows the potential of our approach for shedding light on faint intramolecular dynamics.

Our data also provide a huge amount of new spectroscopic information. In addition to transitions assigned to the CO stretching mode of trioxane, we report line-centre frequencies of more than 400 unassigned transitions most likely belonging to hot bands or to fundamental bands of less abundant isotopologues. Combining frequency metrology and FT spectroscopy can thus contribute to the compilation of spectroscopic data for relatively heavy species (such as atmospheric species SF$_6$, ClONO$_2$, CF$_4$) which with low-lying vibrational modes exhibit, like trioxane, dense rotational structures and many hot bands, impossible to resolve by FT spectroscopy only, and for which the molecular databases thus remain largely incomplete \cite{gordon2022hitran2020}.

Our work demonstrates the potential of our approach for providing new ultra-precise and complete spectroscopic data of larger and larger species, for building accurate spectroscopic models of polyatomic molecules and compiling molecular databases with increasingly more accurate parameters. This is essential for probing fundamental processes in large molecules and identifying the most promising species/transitions amenable to precision measurements for the most demanding tests in fundamental physics, quantum sciences and metrology. Our work for instance unravels a vastly rich rovibrational spectrum of relatively well isolated lines in trioxane which could be used as a grid of references for frequency calibration in the 850-1500 cm$^{-1}$ spectral window, a five times improved spectral coverage compared to the currently used OsO$_4$ \cite{acef1999accurate}, SF$_6$ \cite{acef2000new}, and CO$_2$ \cite{amy2004absolute} absolute frequency grids limited to the 935-1090 cm$^{-1}$ range. 

\section*{SUPPLEMENTARY MATERIAL}
The Supplementary Material provides complementary descriptions of the two ultrahigh-resolution saturated absorption spectroscopic setups respectively using a CO$_2$ laser or a QCL, as well as details on the saturated absorption spectral line shape and line-center frequency determination. Tables listing line-centre frequencies with corresponding uncertainties, assignments, (obs.-calc.) etc. are presented. In addition, additional details are provided for (i) the combined FTIR and saturated absorption data analysis and calibration of the FTIR spectrum, (ii) the comparisons of the results of different Hamiltonian fits, and (iii) the corresponding derived molecular parameters.

\begin{acknowledgments}
We warmly thank Anne Amy-Klein and Andrey Goncharov for their contribution to the development and operation of the QCL-based spectrometer, and to Olivier Lopez for his contributions to all experimental efforts at LPL (and the optical link to LTE ?). We are grateful to Michel Abgrall, Hector Álvarez Martínez (also affiliated and now at Real Instituto y Observatorio de la Armada, San Fernando, Spain) and Rodolphe Le Targat from Laboratoire Temps Espace (LTE, formerly LNE-SYRTE, Observatoire de Paris - PSL, CNRS, Sorbonne Université, LNE) for developing and operating the 1.54 {\textmu}m ultrastable frequency reference calibrated to the primary frequency standards of LTE distributed to the REFIMEVE optical fiber network, and to Etienne Cantin (now at LPL), Dan Xu and Paul-Eric Pottie (LTE) for developing and operating the local links at LTE and the optical fiber link between LTE and LPL. The research leading to these results has received funding from (i) the People Programme (Marie Curie Actions) of the European Union’s Seventh Framework Programme (FP7/2007-2013) under REA grant agreement n. PCOFUND-GA-2013-609102, through the PRESTIGE programme coordinated by Campus France; (ii) the Region Île-de-France in the framework of DIM SIRTEQ; (iii) DRARI and the Region Île-de-France in the framework of CPER Comb’IdF; (iv) the Agence Nationale de la Recherche projects PVCM (Grant No. ANR-15-CE30-0005-01), PSYCHe (Grant No. ANR-17-ERC2-0036-01), DIABOLO (Grant No. ANR-25-CE57-7543) and Ultimos (Grant No. ANR-24-CE30-2959-01), the LabEx Cluster of Excellence FIRST-TF (ANR-10-LABX-48-01) and the EquipEx Cluster of Excellence REFIMEVE+ (ANR-11-EQPX-0039); (v) the European Partnership on Metrology, co-financed from the European Union’s Horizon Europe Research and Innovation Programme and by the Participating States, through the project (23FUN04 COMOMET); (vi) CNRS; (vii) Université Sorbonne Paris Nord. D.B.A. Tran was supported by the Ministry of Education and Training, Vietnam (Program 911). The authors are grateful to synchrotron SOLEIL facility and the infrared beamline AILES staff for giving access to HR Bruker IFS 125 FTIR spectrometer and long path cell under Proposal n° 99240229.
\end{acknowledgments}
\section*{AUTHOR DECLARATIONS}
\subsection*{ Conflict of Interest}
The authors have no conflicts to disclose. 
\subsection*{Author Contributions} 
\noindent \textbf{Pierre Asselin}: conceptualization; methodology; investigation; formal analysis; visualization; data curation; writing - original draft; writing - review and editing. \textbf{Dang-Bao-An Tran}: investigation; formal analysis; visualization; data curation; writing - review and editing. \textbf{Alexander Shelkovnikov}: investigation. \textbf{Sean Tokunaga}: investigation; software; writing - review and editing. \textbf{Mathieu Manceau}: investigation; supervision; funding acquisition; writing - review and editing. \textbf{Benoît Darquié}: conceptualization; methodology; investigation; formal analysis; visualization; data curation; supervision; funding acquisition; project administration; validation; writing - original draft; writing - review and editing.
\section*{Data Availability}
The data that support the findings of this study are available from the corresponding author upon reasonable request.
\section*{References}
\bibliography{aipsamp}

\begin{thebibliography}{43}%
\makeatletter
\providecommand \@ifxundefined [1]{%
 \@ifx{#1\undefined}
}%
\providecommand \@ifnum [1]{%
 \ifnum #1\expandafter \@firstoftwo
 \else \expandafter \@secondoftwo
 \fi
}%
\providecommand \@ifx [1]{%
 \ifx #1\expandafter \@firstoftwo
 \else \expandafter \@secondoftwo
 \fi
}%
\providecommand \natexlab [1]{#1}%
\providecommand \enquote  [1]{``#1''}%
\providecommand \bibnamefont  [1]{#1}%
\providecommand \bibfnamefont [1]{#1}%
\providecommand \citenamefont [1]{#1}%
\providecommand \href@noop [0]{\@secondoftwo}%
\providecommand \href [0]{\begingroup \@sanitize@url \@href}%
\providecommand \@href[1]{\@@startlink{#1}\@@href}%
\providecommand \@@href[1]{\endgroup#1\@@endlink}%
\providecommand \@sanitize@url [0]{\catcode `\\12\catcode `\$12\catcode `\&12\catcode `\#12\catcode `\^12\catcode `\_12\catcode `\%12\relax}%
\providecommand \@@startlink[1]{}%
\providecommand \@@endlink[0]{}%
\providecommand \url  [0]{\begingroup\@sanitize@url \@url }%
\providecommand \@url [1]{\endgroup\@href {#1}{\urlprefix }}%
\providecommand \urlprefix  [0]{URL }%
\providecommand \Eprint [0]{\href }%
\providecommand \doibase [0]{http://dx.doi.org/}%
\providecommand \selectlanguage [0]{\@gobble}%
\providecommand \bibinfo  [0]{\@secondoftwo}%
\providecommand \bibfield  [0]{\@secondoftwo}%
\providecommand \translation [1]{[#1]}%
\providecommand \BibitemOpen [0]{}%
\providecommand \bibitemStop [0]{}%
\providecommand \bibitemNoStop [0]{.\EOS\space}%
\providecommand \EOS [0]{\spacefactor3000\relax}%
\providecommand \BibitemShut  [1]{\csname bibitem#1\endcsname}%
\let\auto@bib@innerbib\@empty
\bibitem [{\citenamefont {Cottin}\ \emph {et~al.}(2000)\citenamefont {Cottin}, \citenamefont {Gazeau}, \citenamefont {Doussin},\ and\ \citenamefont {Raulin}}]{cottin2000experimental}%
  \BibitemOpen
  \bibfield  {author} {\bibinfo {author} {\bibfnamefont {H.}~\bibnamefont {Cottin}}, \bibinfo {author} {\bibfnamefont {M.-C.}\ \bibnamefont {Gazeau}}, \bibinfo {author} {\bibfnamefont {J.-F.}\ \bibnamefont {Doussin}}, \ and\ \bibinfo {author} {\bibfnamefont {F.}~\bibnamefont {Raulin}},\ }\bibfield  {title} {\enquote {\bibinfo {title} {An experimental study of the photodegradation of polyoxymethylene at 122, 147 and 193 nm},}\ }\href@noop {} {\bibfield  {journal} {\bibinfo  {journal} {Journal of photochemistry and photobiology A: Chemistry}\ }\textbf {\bibinfo {volume} {135}},\ \bibinfo {pages} {53--64} (\bibinfo {year} {2000})}\BibitemShut {NoStop}%
\bibitem [{\citenamefont {Cottin}\ \emph {et~al.}(2004)\citenamefont {Cottin}, \citenamefont {B{\'e}nilan}, \citenamefont {Gazeau},\ and\ \citenamefont {Raulin}}]{cottin2004origin}%
  \BibitemOpen
  \bibfield  {author} {\bibinfo {author} {\bibfnamefont {H.}~\bibnamefont {Cottin}}, \bibinfo {author} {\bibfnamefont {Y.}~\bibnamefont {B{\'e}nilan}}, \bibinfo {author} {\bibfnamefont {M.-C.}\ \bibnamefont {Gazeau}}, \ and\ \bibinfo {author} {\bibfnamefont {F.}~\bibnamefont {Raulin}},\ }\bibfield  {title} {\enquote {\bibinfo {title} {Origin of cometary extended sources from degradation of refractory organics on grains: polyoxymethylene as formaldehyde parent molecule},}\ }\href@noop {} {\bibfield  {journal} {\bibinfo  {journal} {Icarus}\ }\textbf {\bibinfo {volume} {167}},\ \bibinfo {pages} {397--416} (\bibinfo {year} {2004})}\BibitemShut {NoStop}%
\bibitem [{\citenamefont {Dangoisse}, \citenamefont {Wascat},\ and\ \citenamefont {Colmont}(1981)}]{dangoisse1981assignment}%
  \BibitemOpen
  \bibfield  {author} {\bibinfo {author} {\bibfnamefont {D.}~\bibnamefont {Dangoisse}}, \bibinfo {author} {\bibfnamefont {J.}~\bibnamefont {Wascat}}, \ and\ \bibinfo {author} {\bibfnamefont {J.}~\bibnamefont {Colmont}},\ }\bibfield  {title} {\enquote {\bibinfo {title} {Assignment of laser lines in an optically pumped submillimeter and near millimeter laser: (\uppercase{H}$_2$\uppercase{CO})$_3$},}\ }\href@noop {} {\bibfield  {journal} {\bibinfo  {journal} {International Journal of Infrared and Millimeter Waves}\ }\textbf {\bibinfo {volume} {2}},\ \bibinfo {pages} {1177--1191} (\bibinfo {year} {1981})}\BibitemShut {NoStop}%
\bibitem [{\citenamefont {Pagies}, \citenamefont {Ducournau},\ and\ \citenamefont {Lampin}(2016)}]{pagies2016low}%
  \BibitemOpen
  \bibfield  {author} {\bibinfo {author} {\bibfnamefont {A.}~\bibnamefont {Pagies}}, \bibinfo {author} {\bibfnamefont {G.}~\bibnamefont {Ducournau}}, \ and\ \bibinfo {author} {\bibfnamefont {J.-F.}\ \bibnamefont {Lampin}},\ }\bibfield  {title} {\enquote {\bibinfo {title} {Low-threshold terahertz molecular laser optically pumped by a quantum cascade laser},}\ }\href@noop {} {\bibfield  {journal} {\bibinfo  {journal} {Apl Photonics}\ }\textbf {\bibinfo {volume} {1}} (\bibinfo {year} {2016})}\BibitemShut {NoStop}%
\bibitem [{\citenamefont {Amirzhan}\ \emph {et~al.}(2022)\citenamefont {Amirzhan}, \citenamefont {Chevalier}, \citenamefont {Rowlette}, \citenamefont {Stinson}, \citenamefont {Pushkarsky}, \citenamefont {Day}, \citenamefont {Everitt},\ and\ \citenamefont {Capasso}}]{amirzhan2022quantum}%
  \BibitemOpen
  \bibfield  {author} {\bibinfo {author} {\bibfnamefont {A.}~\bibnamefont {Amirzhan}}, \bibinfo {author} {\bibfnamefont {P.}~\bibnamefont {Chevalier}}, \bibinfo {author} {\bibfnamefont {J.}~\bibnamefont {Rowlette}}, \bibinfo {author} {\bibfnamefont {H.~T.}\ \bibnamefont {Stinson}}, \bibinfo {author} {\bibfnamefont {M.}~\bibnamefont {Pushkarsky}}, \bibinfo {author} {\bibfnamefont {T.}~\bibnamefont {Day}}, \bibinfo {author} {\bibfnamefont {H.~O.}\ \bibnamefont {Everitt}}, \ and\ \bibinfo {author} {\bibfnamefont {F.}~\bibnamefont {Capasso}},\ }\bibfield  {title} {\enquote {\bibinfo {title} {A quantum cascade laser-pumped molecular laser tunable over 1 \uppercase{TH}z},}\ }\href@noop {} {\bibfield  {journal} {\bibinfo  {journal} {APL Photonics}\ }\textbf {\bibinfo {volume} {7}} (\bibinfo {year} {2022})}\BibitemShut {NoStop}%
\bibitem [{\citenamefont {Acef}, \citenamefont {Michaud},\ and\ \citenamefont {Rovera}(1999)}]{acef1999accurate}%
  \BibitemOpen
  \bibfield  {author} {\bibinfo {author} {\bibfnamefont {O.}~\bibnamefont {Acef}}, \bibinfo {author} {\bibfnamefont {F.}~\bibnamefont {Michaud}}, \ and\ \bibinfo {author} {\bibfnamefont {G.~D.}\ \bibnamefont {Rovera}},\ }\bibfield  {title} {\enquote {\bibinfo {title} {Accurate determination of \uppercase{O}s\uppercase{O}$_4$ absolute frequency grid at 28/29 \uppercase{TH}z},}\ }\href@noop {} {\bibfield  {journal} {\bibinfo  {journal} {IEEE Transactions on Instrumentation and Measurement}\ }\textbf {\bibinfo {volume} {48}},\ \bibinfo {pages} {567--570} (\bibinfo {year} {1999})}\BibitemShut {NoStop}%
\bibitem [{\citenamefont {Acef}\ \emph {et~al.}(2000)\citenamefont {Acef}, \citenamefont {Bord{\'e}}, \citenamefont {Clairon}, \citenamefont {Pierre},\ and\ \citenamefont {Sartakov}}]{acef2000new}%
  \BibitemOpen
  \bibfield  {author} {\bibinfo {author} {\bibfnamefont {O.}~\bibnamefont {Acef}}, \bibinfo {author} {\bibfnamefont {C.~J.}\ \bibnamefont {Bord{\'e}}}, \bibinfo {author} {\bibfnamefont {A.}~\bibnamefont {Clairon}}, \bibinfo {author} {\bibfnamefont {G.}~\bibnamefont {Pierre}}, \ and\ \bibinfo {author} {\bibfnamefont {B.}~\bibnamefont {Sartakov}},\ }\bibfield  {title} {\enquote {\bibinfo {title} {New accurate fit of an extended set of saturation data for the $\nu_3$ band of \uppercase{SF}$_6$: \uppercase{C}omparison of hamiltonians in the spherical and cubic tensor formalisms},}\ }\href@noop {} {\bibfield  {journal} {\bibinfo  {journal} {Journal of Molecular Spectroscopy}\ }\textbf {\bibinfo {volume} {199}},\ \bibinfo {pages} {188--204} (\bibinfo {year} {2000})}\BibitemShut {NoStop}%
\bibitem [{\citenamefont {Amy-Klein}, \citenamefont {Vigu{\'e}},\ and\ \citenamefont {Chardonnet}(2004)}]{amy2004absolute}%
  \BibitemOpen
  \bibfield  {author} {\bibinfo {author} {\bibfnamefont {A.}~\bibnamefont {Amy-Klein}}, \bibinfo {author} {\bibfnamefont {H.}~\bibnamefont {Vigu{\'e}}}, \ and\ \bibinfo {author} {\bibfnamefont {C.}~\bibnamefont {Chardonnet}},\ }\bibfield  {title} {\enquote {\bibinfo {title} {Absolute frequency measurement of $^{12}$\uppercase{C}$^{16}$\uppercase{O}$_2$ laser lines with a femtosecond laser comb and new determination of the $^{12}$\uppercase{C}$^{16}$\uppercase{O}$_2$ molecular constants and frequency grid},}\ }\href@noop {} {\bibfield  {journal} {\bibinfo  {journal} {Journal of Molecular Spectroscopy}\ }\textbf {\bibinfo {volume} {228}},\ \bibinfo {pages} {206--212} (\bibinfo {year} {2004})}\BibitemShut {NoStop}%
\bibitem [{\citenamefont {Oka}\ \emph {et~al.}(1964)\citenamefont {Oka}, \citenamefont {Tsuchiya}, \citenamefont {Iwata},\ and\ \citenamefont {Morino}}]{oka1964microwave}%
  \BibitemOpen
  \bibfield  {author} {\bibinfo {author} {\bibfnamefont {T.}~\bibnamefont {Oka}}, \bibinfo {author} {\bibfnamefont {K.}~\bibnamefont {Tsuchiya}}, \bibinfo {author} {\bibfnamefont {S.}~\bibnamefont {Iwata}}, \ and\ \bibinfo {author} {\bibfnamefont {Y.}~\bibnamefont {Morino}},\ }\bibfield  {title} {\enquote {\bibinfo {title} {Microwave spectrum of s-trioxane},}\ }\href@noop {} {\bibfield  {journal} {\bibinfo  {journal} {Bulletin of the Chemical Society of Japan}\ }\textbf {\bibinfo {volume} {37}},\ \bibinfo {pages} {4--7} (\bibinfo {year} {1964})}\BibitemShut {NoStop}%
\bibitem [{\citenamefont {Colmont}(1975)}]{colmont1975microwave}%
  \BibitemOpen
  \bibfield  {author} {\bibinfo {author} {\bibfnamefont {J.}~\bibnamefont {Colmont}},\ }\bibfield  {title} {\enquote {\bibinfo {title} {Microwave spectrum of trioxane in four excited degenerate vibrational states, especially in the 2$\nu_{20}$ (\uppercase{E}) state and the combination state $\nu_7$ (\uppercase{A}$_1$)+ $\nu_{20}$ (\uppercase{E})},}\ }\href@noop {} {\bibfield  {journal} {\bibinfo  {journal} {Journal of Molecular Spectroscopy}\ }\textbf {\bibinfo {volume} {58}},\ \bibinfo {pages} {220--228} (\bibinfo {year} {1975})}\BibitemShut {NoStop}%
\bibitem [{\citenamefont {Henninot}\ \emph {et~al.}(1992)\citenamefont {Henninot}, \citenamefont {Bolvin}, \citenamefont {Demaison},\ and\ \citenamefont {Lemoine}}]{henninot1992infrared}%
  \BibitemOpen
  \bibfield  {author} {\bibinfo {author} {\bibfnamefont {J.-F.}\ \bibnamefont {Henninot}}, \bibinfo {author} {\bibfnamefont {H.}~\bibnamefont {Bolvin}}, \bibinfo {author} {\bibfnamefont {J.}~\bibnamefont {Demaison}}, \ and\ \bibinfo {author} {\bibfnamefont {B.}~\bibnamefont {Lemoine}},\ }\bibfield  {title} {\enquote {\bibinfo {title} {The infrared spectrum of trioxane in a supersonic slit jet},}\ }\href@noop {} {\bibfield  {journal} {\bibinfo  {journal} {Journal of Molecular Spectroscopy}\ }\textbf {\bibinfo {volume} {152}},\ \bibinfo {pages} {62--68} (\bibinfo {year} {1992})}\BibitemShut {NoStop}%
\bibitem [{\citenamefont {Gibson}, \citenamefont {Koeppen},\ and\ \citenamefont {McCall}(2015)}]{gibson2015rotationally}%
  \BibitemOpen
  \bibfield  {author} {\bibinfo {author} {\bibfnamefont {B.~M.}\ \bibnamefont {Gibson}}, \bibinfo {author} {\bibfnamefont {N.~C.}\ \bibnamefont {Koeppen}}, \ and\ \bibinfo {author} {\bibfnamefont {B.~J.}\ \bibnamefont {McCall}},\ }\bibfield  {title} {\enquote {\bibinfo {title} {Rotationally-resolved spectroscopy of the $\nu_{16}$ band of 1, 3, 5-trioxane},}\ }\href@noop {} {\bibfield  {journal} {\bibinfo  {journal} {Journal of Molecular Spectroscopy}\ }\textbf {\bibinfo {volume} {317}},\ \bibinfo {pages} {47--49} (\bibinfo {year} {2015})}\BibitemShut {NoStop}%
\bibitem [{\citenamefont {Richard}, \citenamefont {Asselin},\ and\ \citenamefont {Boudon}(2022)}]{richard2022high}%
  \BibitemOpen
  \bibfield  {author} {\bibinfo {author} {\bibfnamefont {C.}~\bibnamefont {Richard}}, \bibinfo {author} {\bibfnamefont {P.}~\bibnamefont {Asselin}}, \ and\ \bibinfo {author} {\bibfnamefont {V.}~\bibnamefont {Boudon}},\ }\bibfield  {title} {\enquote {\bibinfo {title} {High-resolution far-infrared synchrotron ftir spectroscopy and analysis of the $\nu_7$, $\nu_{19}$ and $\nu_{20}$ bands of trioxane},}\ }\href@noop {} {\bibfield  {journal} {\bibinfo  {journal} {Journal of Molecular Spectroscopy}\ }\textbf {\bibinfo {volume} {386}},\ \bibinfo {pages} {111614} (\bibinfo {year} {2022})}\BibitemShut {NoStop}%
\bibitem [{\citenamefont {Tran}\ \emph {et~al.}(2025)\citenamefont {Tran}, \citenamefont {Manceau}, \citenamefont {Lopez}, \citenamefont {Goncharov}, \citenamefont {Amy-Klein},\ and\ \citenamefont {Darqui{\'e}}}]{tran2025extending}%
  \BibitemOpen
  \bibfield  {author} {\bibinfo {author} {\bibfnamefont {D.~B.~A.}\ \bibnamefont {Tran}}, \bibinfo {author} {\bibfnamefont {M.}~\bibnamefont {Manceau}}, \bibinfo {author} {\bibfnamefont {O.}~\bibnamefont {Lopez}}, \bibinfo {author} {\bibfnamefont {A.}~\bibnamefont {Goncharov}}, \bibinfo {author} {\bibfnamefont {A.}~\bibnamefont {Amy-Klein}}, \ and\ \bibinfo {author} {\bibfnamefont {B.}~\bibnamefont {Darqui{\'e}}},\ }\bibfield  {title} {\enquote {\bibinfo {title} {Extending frequency metrology to increasingly complex molecules: \uppercase{SI}-traceable sub-\uppercase{D}oppler mid-\uppercase{IR} spectroscopy of trioxane},}\ }\href@noop {} {\bibfield  {journal} {\bibinfo  {journal} {arXiv preprint arXiv:2502.08201}\ } (\bibinfo {year} {2025})}\BibitemShut {NoStop}%
\bibitem [{\citenamefont {Polyansky}\ \emph {et~al.}(2015)\citenamefont {Polyansky}, \citenamefont {Bielska}, \citenamefont {Ghysels}, \citenamefont {Lodi}, \citenamefont {Zobov}, \citenamefont {Hodges},\ and\ \citenamefont {Tennyson}}]{polyansky2015high}%
  \BibitemOpen
  \bibfield  {author} {\bibinfo {author} {\bibfnamefont {O.~L.}\ \bibnamefont {Polyansky}}, \bibinfo {author} {\bibfnamefont {K.}~\bibnamefont {Bielska}}, \bibinfo {author} {\bibfnamefont {M.}~\bibnamefont {Ghysels}}, \bibinfo {author} {\bibfnamefont {L.}~\bibnamefont {Lodi}}, \bibinfo {author} {\bibfnamefont {N.~F.}\ \bibnamefont {Zobov}}, \bibinfo {author} {\bibfnamefont {J.~T.}\ \bibnamefont {Hodges}}, \ and\ \bibinfo {author} {\bibfnamefont {J.}~\bibnamefont {Tennyson}},\ }\bibfield  {title} {\enquote {\bibinfo {title} {High-accuracy \uppercase{CO}$_2$ line intensities determined from theory and experiment},}\ }\href@noop {} {\bibfield  {journal} {\bibinfo  {journal} {Physical review letters}\ }\textbf {\bibinfo {volume} {114}},\ \bibinfo {pages} {243001} (\bibinfo {year} {2015})}\BibitemShut {NoStop}%
\bibitem [{\citenamefont {Vaskuri}\ \emph {et~al.}(2018)\citenamefont {Vaskuri}, \citenamefont {K{\"a}rh{\"a}}, \citenamefont {Egli}, \citenamefont {Gr{\"o}bner},\ and\ \citenamefont {Ikonen}}]{vaskuri2018uncertainty}%
  \BibitemOpen
  \bibfield  {author} {\bibinfo {author} {\bibfnamefont {A.}~\bibnamefont {Vaskuri}}, \bibinfo {author} {\bibfnamefont {P.}~\bibnamefont {K{\"a}rh{\"a}}}, \bibinfo {author} {\bibfnamefont {L.}~\bibnamefont {Egli}}, \bibinfo {author} {\bibfnamefont {J.}~\bibnamefont {Gr{\"o}bner}}, \ and\ \bibinfo {author} {\bibfnamefont {E.}~\bibnamefont {Ikonen}},\ }\bibfield  {title} {\enquote {\bibinfo {title} {Uncertainty analysis of total ozone derived from direct solar irradiance spectra in the presence of unknown spectral deviations},}\ }\href@noop {} {\bibfield  {journal} {\bibinfo  {journal} {Atmospheric Measurement Techniques}\ }\textbf {\bibinfo {volume} {11}},\ \bibinfo {pages} {3595--3610} (\bibinfo {year} {2018})}\BibitemShut {NoStop}%
\bibitem [{\citenamefont {Roueff}\ \emph {et~al.}(2020)\citenamefont {Roueff}, \citenamefont {Sahal-Br{\'e}chot}, \citenamefont {Dimitrijevi{\'c}}, \citenamefont {Moreau},\ and\ \citenamefont {Abgrall}}]{roueff2020spectroscopic}%
  \BibitemOpen
  \bibfield  {author} {\bibinfo {author} {\bibfnamefont {E.}~\bibnamefont {Roueff}}, \bibinfo {author} {\bibfnamefont {S.}~\bibnamefont {Sahal-Br{\'e}chot}}, \bibinfo {author} {\bibfnamefont {M.~S.}\ \bibnamefont {Dimitrijevi{\'c}}}, \bibinfo {author} {\bibfnamefont {N.}~\bibnamefont {Moreau}}, \ and\ \bibinfo {author} {\bibfnamefont {H.}~\bibnamefont {Abgrall}},\ }\bibfield  {title} {\enquote {\bibinfo {title} {The spectroscopic atomic and molecular databases at the paris observatory},}\ }\href@noop {} {\bibfield  {journal} {\bibinfo  {journal} {Atoms}\ }\textbf {\bibinfo {volume} {8}},\ \bibinfo {pages} {36} (\bibinfo {year} {2020})}\BibitemShut {NoStop}%
\bibitem [{\citenamefont {Schulz}\ \emph {et~al.}(2015)\citenamefont {Schulz}, \citenamefont {Ij{\"a}s}, \citenamefont {Drost}, \citenamefont {H{\"a}m{\"a}l{\"a}inen}, \citenamefont {Harju}, \citenamefont {Seitsonen},\ and\ \citenamefont {Liljeroth}}]{schulz2015many}%
  \BibitemOpen
  \bibfield  {author} {\bibinfo {author} {\bibfnamefont {F.}~\bibnamefont {Schulz}}, \bibinfo {author} {\bibfnamefont {M.}~\bibnamefont {Ij{\"a}s}}, \bibinfo {author} {\bibfnamefont {R.}~\bibnamefont {Drost}}, \bibinfo {author} {\bibfnamefont {S.~K.}\ \bibnamefont {H{\"a}m{\"a}l{\"a}inen}}, \bibinfo {author} {\bibfnamefont {A.}~\bibnamefont {Harju}}, \bibinfo {author} {\bibfnamefont {A.~P.}\ \bibnamefont {Seitsonen}}, \ and\ \bibinfo {author} {\bibfnamefont {P.}~\bibnamefont {Liljeroth}},\ }\bibfield  {title} {\enquote {\bibinfo {title} {Many-body transitions in a single molecule visualized by scanning tunnelling microscopy},}\ }\href@noop {} {\bibfield  {journal} {\bibinfo  {journal} {Nature Physics}\ }\textbf {\bibinfo {volume} {11}},\ \bibinfo {pages} {229--234} (\bibinfo {year} {2015})}\BibitemShut {NoStop}%
\bibitem [{\citenamefont {Liu}\ \emph {et~al.}(2023)\citenamefont {Liu}, \citenamefont {Rosenberg}, \citenamefont {Changala}, \citenamefont {Crowley}, \citenamefont {Nesbitt}, \citenamefont {Yao}, \citenamefont {Tscherbul},\ and\ \citenamefont {Ye}}]{liu2023ergodicity}%
  \BibitemOpen
  \bibfield  {author} {\bibinfo {author} {\bibfnamefont {L.~R.}\ \bibnamefont {Liu}}, \bibinfo {author} {\bibfnamefont {D.}~\bibnamefont {Rosenberg}}, \bibinfo {author} {\bibfnamefont {P.~B.}\ \bibnamefont {Changala}}, \bibinfo {author} {\bibfnamefont {P.~J.}\ \bibnamefont {Crowley}}, \bibinfo {author} {\bibfnamefont {D.~J.}\ \bibnamefont {Nesbitt}}, \bibinfo {author} {\bibfnamefont {N.~Y.}\ \bibnamefont {Yao}}, \bibinfo {author} {\bibfnamefont {T.~V.}\ \bibnamefont {Tscherbul}}, \ and\ \bibinfo {author} {\bibfnamefont {J.}~\bibnamefont {Ye}},\ }\bibfield  {title} {\enquote {\bibinfo {title} {Ergodicity breaking in rapidly rotating \uppercase{C}$_{60}$ fullerenes},}\ }\href@noop {} {\bibfield  {journal} {\bibinfo  {journal} {Science}\ }\textbf {\bibinfo {volume} {381}},\ \bibinfo {pages} {778--783} (\bibinfo {year} {2023})}\BibitemShut {NoStop}%
\bibitem [{\citenamefont {Tesch}\ and\ \citenamefont {de~Vivie-Riedle}(2002)}]{tesch2002quantum}%
  \BibitemOpen
  \bibfield  {author} {\bibinfo {author} {\bibfnamefont {C.~M.}\ \bibnamefont {Tesch}}\ and\ \bibinfo {author} {\bibfnamefont {R.}~\bibnamefont {de~Vivie-Riedle}},\ }\bibfield  {title} {\enquote {\bibinfo {title} {Quantum computation with vibrationally excited molecules},}\ }\href@noop {} {\bibfield  {journal} {\bibinfo  {journal} {Physical review letters}\ }\textbf {\bibinfo {volume} {89}},\ \bibinfo {pages} {157901} (\bibinfo {year} {2002})}\BibitemShut {NoStop}%
\bibitem [{\citenamefont {Albert}\ \emph {et~al.}(2024)\citenamefont {Albert}, \citenamefont {Kubischta}, \citenamefont {Lemeshko},\ and\ \citenamefont {Liu}}]{albert2024quantum}%
  \BibitemOpen
  \bibfield  {author} {\bibinfo {author} {\bibfnamefont {V.~V.}\ \bibnamefont {Albert}}, \bibinfo {author} {\bibfnamefont {E.}~\bibnamefont {Kubischta}}, \bibinfo {author} {\bibfnamefont {M.}~\bibnamefont {Lemeshko}}, \ and\ \bibinfo {author} {\bibfnamefont {L.~R.}\ \bibnamefont {Liu}},\ }\bibfield  {title} {\enquote {\bibinfo {title} {Quantum theory of molecular orientations},}\ }\href@noop {} {\bibfield  {journal} {\bibinfo  {journal} {arXiv preprint arXiv:2403.04572}\ } (\bibinfo {year} {2024})}\BibitemShut {NoStop}%
\bibitem [{\citenamefont {Shelkovnikov}\ \emph {et~al.}(2008)\citenamefont {Shelkovnikov}, \citenamefont {Butcher}, \citenamefont {Chardonnet},\ and\ \citenamefont {Amy-Klein}}]{shelkovnikov2008stability}%
  \BibitemOpen
  \bibfield  {author} {\bibinfo {author} {\bibfnamefont {A.}~\bibnamefont {Shelkovnikov}}, \bibinfo {author} {\bibfnamefont {R.~J.}\ \bibnamefont {Butcher}}, \bibinfo {author} {\bibfnamefont {C.}~\bibnamefont {Chardonnet}}, \ and\ \bibinfo {author} {\bibfnamefont {A.}~\bibnamefont {Amy-Klein}},\ }\bibfield  {title} {\enquote {\bibinfo {title} {Stability of the proton-to-electron mass ratio},}\ }\href@noop {} {\bibfield  {journal} {\bibinfo  {journal} {Physical Review Letters}\ }\textbf {\bibinfo {volume} {100}},\ \bibinfo {pages} {150801} (\bibinfo {year} {2008})}\BibitemShut {NoStop}%
\bibitem [{\citenamefont {Jansen}, \citenamefont {Bethlem},\ and\ \citenamefont {Ubachs}(2014)}]{jansen2014perspective}%
  \BibitemOpen
  \bibfield  {author} {\bibinfo {author} {\bibfnamefont {P.}~\bibnamefont {Jansen}}, \bibinfo {author} {\bibfnamefont {H.~L.}\ \bibnamefont {Bethlem}}, \ and\ \bibinfo {author} {\bibfnamefont {W.}~\bibnamefont {Ubachs}},\ }\bibfield  {title} {\enquote {\bibinfo {title} {Perspective: \uppercase{T}ipping the scales: \uppercase{S}earch for drifting constants from molecular spectra},}\ }\href@noop {} {\bibfield  {journal} {\bibinfo  {journal} {The Journal of chemical physics}\ }\textbf {\bibinfo {volume} {140}} (\bibinfo {year} {2014})}\BibitemShut {NoStop}%
\bibitem [{\citenamefont {Muller}\ \emph {et~al.}(2021)\citenamefont {Muller}, \citenamefont {Ubachs}, \citenamefont {Menten}, \citenamefont {Henkel},\ and\ \citenamefont {Kanekar}}]{muller2021study}%
  \BibitemOpen
  \bibfield  {author} {\bibinfo {author} {\bibfnamefont {S.}~\bibnamefont {Muller}}, \bibinfo {author} {\bibfnamefont {W.}~\bibnamefont {Ubachs}}, \bibinfo {author} {\bibfnamefont {K.}~\bibnamefont {Menten}}, \bibinfo {author} {\bibfnamefont {C.}~\bibnamefont {Henkel}}, \ and\ \bibinfo {author} {\bibfnamefont {N.}~\bibnamefont {Kanekar}},\ }\bibfield  {title} {\enquote {\bibinfo {title} {A study of submillimeter methanol absorption toward \uppercase{PKS} 1830- 211:-\uppercase{E}xcitation, invariance of the proton-electron mass ratio, and systematics},}\ }\href@noop {} {\bibfield  {journal} {\bibinfo  {journal} {Astronomy \& Astrophysics}\ }\textbf {\bibinfo {volume} {652}},\ \bibinfo {pages} {A5} (\bibinfo {year} {2021})}\BibitemShut {NoStop}%
\bibitem [{\citenamefont {Fiechter}\ \emph {et~al.}(2022)\citenamefont {Fiechter}, \citenamefont {Haase}, \citenamefont {Saleh}, \citenamefont {Soulard}, \citenamefont {Tremblay}, \citenamefont {Havenith}, \citenamefont {Timmermans}, \citenamefont {Schwerdtfeger}, \citenamefont {Crassous}, \citenamefont {Darqui{\'e}} \emph {et~al.}}]{fiechter2022toward}%
  \BibitemOpen
  \bibfield  {author} {\bibinfo {author} {\bibfnamefont {M.~R.}\ \bibnamefont {Fiechter}}, \bibinfo {author} {\bibfnamefont {P.~A.}\ \bibnamefont {Haase}}, \bibinfo {author} {\bibfnamefont {N.}~\bibnamefont {Saleh}}, \bibinfo {author} {\bibfnamefont {P.}~\bibnamefont {Soulard}}, \bibinfo {author} {\bibfnamefont {B.}~\bibnamefont {Tremblay}}, \bibinfo {author} {\bibfnamefont {R.~W.}\ \bibnamefont {Havenith}}, \bibinfo {author} {\bibfnamefont {R.~G.}\ \bibnamefont {Timmermans}}, \bibinfo {author} {\bibfnamefont {P.}~\bibnamefont {Schwerdtfeger}}, \bibinfo {author} {\bibfnamefont {J.}~\bibnamefont {Crassous}}, \bibinfo {author} {\bibfnamefont {B.}~\bibnamefont {Darqui{\'e}}},  \emph {et~al.},\ }\bibfield  {title} {\enquote {\bibinfo {title} {Toward detection of the molecular parity violation in chiral \uppercase{R}u(acac)$_3$ and \uppercase{O}s(acac)$_3$},}\ }\href@noop {} {\bibfield  {journal} {\bibinfo  {journal} {The Journal of Physical Chemistry Letters}\ }\textbf {\bibinfo {volume} {13}},\ \bibinfo {pages}
  {10011--10017} (\bibinfo {year} {2022})}\BibitemShut {NoStop}%
\bibitem [{\citenamefont {Gaul}\ \emph {et~al.}(2020)\citenamefont {Gaul}, \citenamefont {Kozlov}, \citenamefont {Isaev},\ and\ \citenamefont {Berger}}]{gaul2020chiral}%
  \BibitemOpen
  \bibfield  {author} {\bibinfo {author} {\bibfnamefont {K.}~\bibnamefont {Gaul}}, \bibinfo {author} {\bibfnamefont {M.~G.}\ \bibnamefont {Kozlov}}, \bibinfo {author} {\bibfnamefont {T.~A.}\ \bibnamefont {Isaev}}, \ and\ \bibinfo {author} {\bibfnamefont {R.}~\bibnamefont {Berger}},\ }\bibfield  {title} {\enquote {\bibinfo {title} {Chiral molecules as sensitive probes for direct detection of \uppercase{P}-odd cosmic fields},}\ }\href@noop {} {\bibfield  {journal} {\bibinfo  {journal} {Physical review letters}\ }\textbf {\bibinfo {volume} {125}},\ \bibinfo {pages} {123004} (\bibinfo {year} {2020})}\BibitemShut {NoStop}%
\bibitem [{\citenamefont {De~Wit}, \citenamefont {Van~Miltenburg},\ and\ \citenamefont {De~Kruif}(1983)}]{de1983thermodynamic}%
  \BibitemOpen
  \bibfield  {author} {\bibinfo {author} {\bibfnamefont {H.}~\bibnamefont {De~Wit}}, \bibinfo {author} {\bibfnamefont {J.}~\bibnamefont {Van~Miltenburg}}, \ and\ \bibinfo {author} {\bibfnamefont {C.}~\bibnamefont {De~Kruif}},\ }\bibfield  {title} {\enquote {\bibinfo {title} {Thermodynamic properties of molecular organic crystals containing nitrogen, oxygen, and sulphur 1. \uppercase{V}apour pressures and enthalpies of sublimation},}\ }\href@noop {} {\bibfield  {journal} {\bibinfo  {journal} {The Journal of Chemical Thermodynamics}\ }\textbf {\bibinfo {volume} {15}},\ \bibinfo {pages} {651--663} (\bibinfo {year} {1983})}\BibitemShut {NoStop}%
\bibitem [{\citenamefont {Gordon}\ \emph {et~al.}(2022)\citenamefont {Gordon}, \citenamefont {Rothman}, \citenamefont {Hargreaves}, \citenamefont {Hashemi}, \citenamefont {Karlovets}, \citenamefont {Skinner}, \citenamefont {Conway}, \citenamefont {Hill}, \citenamefont {Kochanov}, \citenamefont {Tan} \emph {et~al.}}]{gordon2022hitran2020}%
  \BibitemOpen
  \bibfield  {author} {\bibinfo {author} {\bibfnamefont {I.~E.}\ \bibnamefont {Gordon}}, \bibinfo {author} {\bibfnamefont {L.~S.}\ \bibnamefont {Rothman}}, \bibinfo {author} {\bibfnamefont {e.~R.}\ \bibnamefont {Hargreaves}}, \bibinfo {author} {\bibfnamefont {R.}~\bibnamefont {Hashemi}}, \bibinfo {author} {\bibfnamefont {E.~V.}\ \bibnamefont {Karlovets}}, \bibinfo {author} {\bibfnamefont {F.}~\bibnamefont {Skinner}}, \bibinfo {author} {\bibfnamefont {E.~K.}\ \bibnamefont {Conway}}, \bibinfo {author} {\bibfnamefont {C.}~\bibnamefont {Hill}}, \bibinfo {author} {\bibfnamefont {R.~V.}\ \bibnamefont {Kochanov}}, \bibinfo {author} {\bibfnamefont {Y.}~\bibnamefont {Tan}},  \emph {et~al.},\ }\bibfield  {title} {\enquote {\bibinfo {title} {The \uppercase{HITRAN2020} molecular spectroscopic database},}\ }\href@noop {} {\bibfield  {journal} {\bibinfo  {journal} {Journal of quantitative spectroscopy and radiative transfer}\ }\textbf {\bibinfo {volume} {277}},\ \bibinfo {pages} {107949} (\bibinfo {year}
  {2022})}\BibitemShut {NoStop}%
\bibitem [{\citenamefont {Stoeffler}\ \emph {et~al.}(2011)\citenamefont {Stoeffler}, \citenamefont {Darqui{\'e}}, \citenamefont {Shelkovnikov}, \citenamefont {Daussy}, \citenamefont {Amy-Klein}, \citenamefont {Chardonnet}, \citenamefont {Guy}, \citenamefont {Crassous}, \citenamefont {Huet}, \citenamefont {Soulard} \emph {et~al.}}]{stoeffler2011high}%
  \BibitemOpen
  \bibfield  {author} {\bibinfo {author} {\bibfnamefont {C.}~\bibnamefont {Stoeffler}}, \bibinfo {author} {\bibfnamefont {B.}~\bibnamefont {Darqui{\'e}}}, \bibinfo {author} {\bibfnamefont {A.}~\bibnamefont {Shelkovnikov}}, \bibinfo {author} {\bibfnamefont {C.}~\bibnamefont {Daussy}}, \bibinfo {author} {\bibfnamefont {A.}~\bibnamefont {Amy-Klein}}, \bibinfo {author} {\bibfnamefont {C.}~\bibnamefont {Chardonnet}}, \bibinfo {author} {\bibfnamefont {L.}~\bibnamefont {Guy}}, \bibinfo {author} {\bibfnamefont {J.}~\bibnamefont {Crassous}}, \bibinfo {author} {\bibfnamefont {T.~R.}\ \bibnamefont {Huet}}, \bibinfo {author} {\bibfnamefont {P.}~\bibnamefont {Soulard}},  \emph {et~al.},\ }\bibfield  {title} {\enquote {\bibinfo {title} {High resolution spectroscopy of methyltrioxorhenium: towards the observation of parity violation in chiral molecules},}\ }\href@noop {} {\bibfield  {journal} {\bibinfo  {journal} {Physical Chemistry Chemical Physics}\ }\textbf {\bibinfo {volume} {13}},\ \bibinfo {pages} {854--863} (\bibinfo
  {year} {2011})}\BibitemShut {NoStop}%
\bibitem [{\citenamefont {Argence}\ \emph {et~al.}(2015)\citenamefont {Argence}, \citenamefont {Chanteau}, \citenamefont {Lopez}, \citenamefont {Nicolodi}, \citenamefont {Abgrall}, \citenamefont {Chardonnet}, \citenamefont {Daussy}, \citenamefont {Darqui{\'e}}, \citenamefont {Le~Coq},\ and\ \citenamefont {Amy-Klein}}]{argence2015quantum}%
  \BibitemOpen
  \bibfield  {author} {\bibinfo {author} {\bibfnamefont {B.}~\bibnamefont {Argence}}, \bibinfo {author} {\bibfnamefont {B.}~\bibnamefont {Chanteau}}, \bibinfo {author} {\bibfnamefont {O.}~\bibnamefont {Lopez}}, \bibinfo {author} {\bibfnamefont {D.}~\bibnamefont {Nicolodi}}, \bibinfo {author} {\bibfnamefont {M.}~\bibnamefont {Abgrall}}, \bibinfo {author} {\bibfnamefont {C.}~\bibnamefont {Chardonnet}}, \bibinfo {author} {\bibfnamefont {C.}~\bibnamefont {Daussy}}, \bibinfo {author} {\bibfnamefont {B.}~\bibnamefont {Darqui{\'e}}}, \bibinfo {author} {\bibfnamefont {Y.}~\bibnamefont {Le~Coq}}, \ and\ \bibinfo {author} {\bibfnamefont {A.}~\bibnamefont {Amy-Klein}},\ }\bibfield  {title} {\enquote {\bibinfo {title} {Quantum cascade laser frequency stabilization at the sub-\uppercase{H}z level},}\ }\href@noop {} {\bibfield  {journal} {\bibinfo  {journal} {Nature Photonics}\ }\textbf {\bibinfo {volume} {9}},\ \bibinfo {pages} {456--460} (\bibinfo {year} {2015})}\BibitemShut {NoStop}%
\bibitem [{\citenamefont {Santagata}\ \emph {et~al.}(2019)\citenamefont {Santagata}, \citenamefont {Tran}, \citenamefont {Argence}, \citenamefont {Lopez}, \citenamefont {Tokunaga}, \citenamefont {Wiotte}, \citenamefont {Mouhamad}, \citenamefont {Goncharov}, \citenamefont {Abgrall}, \citenamefont {Le~Coq} \emph {et~al.}}]{santagata2019high}%
  \BibitemOpen
  \bibfield  {author} {\bibinfo {author} {\bibfnamefont {R.}~\bibnamefont {Santagata}}, \bibinfo {author} {\bibfnamefont {D.~B.~A.}\ \bibnamefont {Tran}}, \bibinfo {author} {\bibfnamefont {B.}~\bibnamefont {Argence}}, \bibinfo {author} {\bibfnamefont {O.}~\bibnamefont {Lopez}}, \bibinfo {author} {\bibfnamefont {S.}~\bibnamefont {Tokunaga}}, \bibinfo {author} {\bibfnamefont {F.}~\bibnamefont {Wiotte}}, \bibinfo {author} {\bibfnamefont {H.}~\bibnamefont {Mouhamad}}, \bibinfo {author} {\bibfnamefont {A.}~\bibnamefont {Goncharov}}, \bibinfo {author} {\bibfnamefont {M.}~\bibnamefont {Abgrall}}, \bibinfo {author} {\bibfnamefont {Y.}~\bibnamefont {Le~Coq}},  \emph {et~al.},\ }\bibfield  {title} {\enquote {\bibinfo {title} {High-precision methanol spectroscopy with a widely tunable \uppercase{SI}-traceable frequency-comb-based mid-infrared \uppercase{QCL}},}\ }\href@noop {} {\bibfield  {journal} {\bibinfo  {journal} {Optica}\ }\textbf {\bibinfo {volume} {6}},\ \bibinfo {pages} {411--423} (\bibinfo {year}
  {2019})}\BibitemShut {NoStop}%
\bibitem [{\citenamefont {Tran}\ \emph {et~al.}(2024)\citenamefont {Tran}, \citenamefont {Lopez}, \citenamefont {Manceau}, \citenamefont {Goncharov}, \citenamefont {Abgrall}, \citenamefont {Alvarez-Martinez}, \citenamefont {Le~Targat}, \citenamefont {Cantin}, \citenamefont {Pottie}, \citenamefont {Amy-Klein} \emph {et~al.}}]{tran2024near}%
  \BibitemOpen
  \bibfield  {author} {\bibinfo {author} {\bibfnamefont {D.~B.~A.}\ \bibnamefont {Tran}}, \bibinfo {author} {\bibfnamefont {O.}~\bibnamefont {Lopez}}, \bibinfo {author} {\bibfnamefont {M.}~\bibnamefont {Manceau}}, \bibinfo {author} {\bibfnamefont {A.}~\bibnamefont {Goncharov}}, \bibinfo {author} {\bibfnamefont {M.}~\bibnamefont {Abgrall}}, \bibinfo {author} {\bibfnamefont {H.}~\bibnamefont {Alvarez-Martinez}}, \bibinfo {author} {\bibfnamefont {R.}~\bibnamefont {Le~Targat}}, \bibinfo {author} {\bibfnamefont {E.}~\bibnamefont {Cantin}}, \bibinfo {author} {\bibfnamefont {P.-E.}\ \bibnamefont {Pottie}}, \bibinfo {author} {\bibfnamefont {A.}~\bibnamefont {Amy-Klein}},  \emph {et~al.},\ }\bibfield  {title} {\enquote {\bibinfo {title} {Near-to mid-\uppercase{IR} spectral purity transfer with a tunable frequency comb: \uppercase{M}ethanol frequency metrology over a 1.4 \uppercase{GH}z span},}\ }\href@noop {} {\bibfield  {journal} {\bibinfo  {journal} {APL Photonics}\ }\textbf {\bibinfo {volume} {9}} (\bibinfo {year}
  {2024})}\BibitemShut {NoStop}%
\bibitem [{\citenamefont {Cantin}\ \emph {et~al.}(2021)\citenamefont {Cantin}, \citenamefont {T{\o}nnes}, \citenamefont {Le~Targat}, \citenamefont {Amy-Klein}, \citenamefont {Lopez},\ and\ \citenamefont {Pottie}}]{cantin2021accurate}%
  \BibitemOpen
  \bibfield  {author} {\bibinfo {author} {\bibfnamefont {E.}~\bibnamefont {Cantin}}, \bibinfo {author} {\bibfnamefont {M.}~\bibnamefont {T{\o}nnes}}, \bibinfo {author} {\bibfnamefont {R.}~\bibnamefont {Le~Targat}}, \bibinfo {author} {\bibfnamefont {A.}~\bibnamefont {Amy-Klein}}, \bibinfo {author} {\bibfnamefont {O.}~\bibnamefont {Lopez}}, \ and\ \bibinfo {author} {\bibfnamefont {P.-E.}\ \bibnamefont {Pottie}},\ }\bibfield  {title} {\enquote {\bibinfo {title} {An accurate and robust metrological network for coherent optical frequency dissemination},}\ }\href@noop {} {\bibfield  {journal} {\bibinfo  {journal} {New Journal of Physics}\ }\textbf {\bibinfo {volume} {23}},\ \bibinfo {pages} {053027} (\bibinfo {year} {2021})}\BibitemShut {NoStop}%
\bibitem [{\citenamefont {Kobayashi}, \citenamefont {Iwamoto},\ and\ \citenamefont {Tadokoro}(1966)}]{kobayashi1966vibrational}%
  \BibitemOpen
  \bibfield  {author} {\bibinfo {author} {\bibfnamefont {M.}~\bibnamefont {Kobayashi}}, \bibinfo {author} {\bibfnamefont {R.}~\bibnamefont {Iwamoto}}, \ and\ \bibinfo {author} {\bibfnamefont {H.}~\bibnamefont {Tadokoro}},\ }\bibfield  {title} {\enquote {\bibinfo {title} {Vibrational spectra of \uppercase{T}rioxane and \uppercase{T}rioxane-$d_6$},}\ }\href@noop {} {\bibfield  {journal} {\bibinfo  {journal} {The Journal of Chemical Physics}\ }\textbf {\bibinfo {volume} {44}},\ \bibinfo {pages} {922--933} (\bibinfo {year} {1966})}\BibitemShut {NoStop}%
\bibitem [{\citenamefont {Klein}, \citenamefont {Belov},\ and\ \citenamefont {Winnewisser}(1996)}]{klein1996terahertz}%
  \BibitemOpen
  \bibfield  {author} {\bibinfo {author} {\bibfnamefont {H.}~\bibnamefont {Klein}}, \bibinfo {author} {\bibfnamefont {S.}~\bibnamefont {Belov}}, \ and\ \bibinfo {author} {\bibfnamefont {G.}~\bibnamefont {Winnewisser}},\ }\bibfield  {title} {\enquote {\bibinfo {title} {Terahertz spectrum of trioxane},}\ }\href@noop {} {\bibfield  {journal} {\bibinfo  {journal} {Zeitschrift f{\"u}r Naturforschung A}\ }\textbf {\bibinfo {volume} {51}},\ \bibinfo {pages} {123--128} (\bibinfo {year} {1996})}\BibitemShut {NoStop}%
\bibitem [{\citenamefont {Rohart}\ \emph {et~al.}(2014)\citenamefont {Rohart}, \citenamefont {Mejri}, \citenamefont {Sow}, \citenamefont {Tokunaga}, \citenamefont {Chardonnet}, \citenamefont {Darqui{\'e}}, \citenamefont {Dinesan}, \citenamefont {Fasci}, \citenamefont {Castrillo}, \citenamefont {Gianfrani} \emph {et~al.}}]{rohart2014absorption}%
  \BibitemOpen
  \bibfield  {author} {\bibinfo {author} {\bibfnamefont {F.}~\bibnamefont {Rohart}}, \bibinfo {author} {\bibfnamefont {S.}~\bibnamefont {Mejri}}, \bibinfo {author} {\bibfnamefont {P.~L.~T.}\ \bibnamefont {Sow}}, \bibinfo {author} {\bibfnamefont {S.~K.}\ \bibnamefont {Tokunaga}}, \bibinfo {author} {\bibfnamefont {C.}~\bibnamefont {Chardonnet}}, \bibinfo {author} {\bibfnamefont {B.}~\bibnamefont {Darqui{\'e}}}, \bibinfo {author} {\bibfnamefont {H.}~\bibnamefont {Dinesan}}, \bibinfo {author} {\bibfnamefont {E.}~\bibnamefont {Fasci}}, \bibinfo {author} {\bibfnamefont {A.}~\bibnamefont {Castrillo}}, \bibinfo {author} {\bibfnamefont {L.}~\bibnamefont {Gianfrani}},  \emph {et~al.},\ }\bibfield  {title} {\enquote {\bibinfo {title} {Absorption-line-shape recovery beyond the detection-bandwidth limit: \uppercase{A}pplication to the precision spectroscopic measurement of the \uppercase{B}oltzmann constant},}\ }\href@noop {} {\bibfield  {journal} {\bibinfo  {journal} {Physical Review A}\ }\textbf {\bibinfo {volume}
  {90}},\ \bibinfo {pages} {042506} (\bibinfo {year} {2014})}\BibitemShut {NoStop}%
\bibitem [{\citenamefont {Rohart}(2017)}]{rohart2017overcoming}%
  \BibitemOpen
  \bibfield  {author} {\bibinfo {author} {\bibfnamefont {F.}~\bibnamefont {Rohart}},\ }\bibfield  {title} {\enquote {\bibinfo {title} {Overcoming the detection bandwidth limit in precision spectroscopy: \uppercase{T}he analytical apparatus function for a stepped frequency scan},}\ }\href@noop {} {\bibfield  {journal} {\bibinfo  {journal} {Journal of Quantitative Spectroscopy and Radiative Transfer}\ }\textbf {\bibinfo {volume} {187}},\ \bibinfo {pages} {490--504} (\bibinfo {year} {2017})}\BibitemShut {NoStop}%
\bibitem [{\citenamefont {Arndt}(1965)}]{arndt1965analytical}%
  \BibitemOpen
  \bibfield  {author} {\bibinfo {author} {\bibfnamefont {R.}~\bibnamefont {Arndt}},\ }\bibfield  {title} {\enquote {\bibinfo {title} {Analytical line shapes for lorentzian signals broadened by modulation},}\ }\href@noop {} {\bibfield  {journal} {\bibinfo  {journal} {Journal of Applied Physics}\ }\textbf {\bibinfo {volume} {36}},\ \bibinfo {pages} {2522--2524} (\bibinfo {year} {1965})}\BibitemShut {NoStop}%
\bibitem [{\citenamefont {Schilt}, \citenamefont {Thevenaz},\ and\ \citenamefont {Robert}(2003)}]{schilt2003wavelength}%
  \BibitemOpen
  \bibfield  {author} {\bibinfo {author} {\bibfnamefont {S.}~\bibnamefont {Schilt}}, \bibinfo {author} {\bibfnamefont {L.}~\bibnamefont {Thevenaz}}, \ and\ \bibinfo {author} {\bibfnamefont {P.}~\bibnamefont {Robert}},\ }\bibfield  {title} {\enquote {\bibinfo {title} {Wavelength modulation spectroscopy: combined frequency and intensity laser modulation},}\ }\href@noop {} {\bibfield  {journal} {\bibinfo  {journal} {Applied optics}\ }\textbf {\bibinfo {volume} {42}},\ \bibinfo {pages} {6728--6738} (\bibinfo {year} {2003})}\BibitemShut {NoStop}%
\bibitem [{\citenamefont {Western}(2017)}]{western2017pgopher}%
  \BibitemOpen
  \bibfield  {author} {\bibinfo {author} {\bibfnamefont {C.~M.}\ \bibnamefont {Western}},\ }\bibfield  {title} {\enquote {\bibinfo {title} {\uppercase{PGOPHER}: \uppercase{A} program for simulating rotational, vibrational and electronic spectra},}\ }\href@noop {} {\bibfield  {journal} {\bibinfo  {journal} {Journal of Quantitative Spectroscopy and Radiative Transfer}\ }\textbf {\bibinfo {volume} {186}},\ \bibinfo {pages} {221--242} (\bibinfo {year} {2017})}\BibitemShut {NoStop}%
\bibitem [{\citenamefont {Pickett}(1991)}]{pickett1991fitting}%
  \BibitemOpen
  \bibfield  {author} {\bibinfo {author} {\bibfnamefont {H.~M.}\ \bibnamefont {Pickett}},\ }\bibfield  {title} {\enquote {\bibinfo {title} {The fitting and prediction of vibration-rotation spectra with spin interactions},}\ }\href@noop {} {\bibfield  {journal} {\bibinfo  {journal} {Journal of Molecular spectroscopy}\ }\textbf {\bibinfo {volume} {148}},\ \bibinfo {pages} {371--377} (\bibinfo {year} {1991})}\BibitemShut {NoStop}%
\bibitem [{\citenamefont {Gadhi}\ \emph {et~al.}(1989)\citenamefont {Gadhi}, \citenamefont {Wlodarczak}, \citenamefont {Boucher},\ and\ \citenamefont {Demaison}}]{gadhi1989submillimeter}%
  \BibitemOpen
  \bibfield  {author} {\bibinfo {author} {\bibfnamefont {J.}~\bibnamefont {Gadhi}}, \bibinfo {author} {\bibfnamefont {G.}~\bibnamefont {Wlodarczak}}, \bibinfo {author} {\bibfnamefont {D.}~\bibnamefont {Boucher}}, \ and\ \bibinfo {author} {\bibfnamefont {J.}~\bibnamefont {Demaison}},\ }\bibfield  {title} {\enquote {\bibinfo {title} {The submillimeter-wave spectrum of trioxane},}\ }\href@noop {} {\bibfield  {journal} {\bibinfo  {journal} {Journal of Molecular Spectroscopy}\ }\textbf {\bibinfo {volume} {133}},\ \bibinfo {pages} {406--412} (\bibinfo {year} {1989})}\BibitemShut {NoStop}%
\bibitem [{\citenamefont {Nesbitt}\ and\ \citenamefont {Field}(1996)}]{nesbitt1996vibrational}%
  \BibitemOpen
  \bibfield  {author} {\bibinfo {author} {\bibfnamefont {D.~J.}\ \bibnamefont {Nesbitt}}\ and\ \bibinfo {author} {\bibfnamefont {R.~W.}\ \bibnamefont {Field}},\ }\bibfield  {title} {\enquote {\bibinfo {title} {Vibrational energy flow in highly excited molecules: Role of intramolecular vibrational redistribution},}\ }\href@noop {} {\bibfield  {journal} {\bibinfo  {journal} {The Journal of Physical Chemistry}\ }\textbf {\bibinfo {volume} {100}},\ \bibinfo {pages} {12735--12756} (\bibinfo {year} {1996})}\BibitemShut {NoStop}%
\end{thebibliography}%

\end{document}